\documentclass{aastex}
\begin{document}

\title{Spectral Irradiance Calibration in the Infrared. XIII. ``Supertemplates"
and On-Orbit Calibrators for SIRTF's Infrared Array Camera (IRAC)}

\author{Martin Cohen}
\affil{Radio Astronomy Laboratory, 601 Campbell Hall, University of
California, Berkeley, California 94720\\
Electronic Mail: mcohen@astro.berkeley.edu}

\author{S. T. Megeath}
\affil{Harvard-Smithsonian Center for Astrophysics, 60 Garden Street, Cambridge, MA 02138\\
Electronic Mail: tmegeath@cfa.harvard.edu}

\author{Peter L. Hammersley}
\affil{Instituto de Astrof\'{\i}sica de Canarias, La Laguna, Tenerife\\
Electronic Mail: plh@ll.iac.es}

\author{Fabiola Martin-Luis}
\affil{Instituto de Astrof\'{\i}sica de Canarias, La Laguna, Tenerife\\
Electronic Mail: fabi@ll.iac.es}

\author{John Stauffer}
\affil{California Institute of Technology, SIRTF Science Center, MS 314-6,
Pasadena, CA 91125\\
Electronic Mail: stauffer@ipac.caltech.edu}

\begin{abstract}
We describe the technique that will be used to develop a set of 
on-orbit calibrators for IRAC and demonstrate the validity of the method
for stars with spectral types either K0-M0III or A0-A5V.  
For application to SIRTF, the approach is intended to 
operate with all available optical, near-infrared (NIR), and mid-infrared (MIR)
photometry, and to yield complete absolute spectra from UV to MIR.  One set of 
stars is picked from Landolt's extensive network of optical ($UBVRI$) 
calibrators; the other from the Carter-Meadows set of faint infrared 
standards.  Traceability to the ``Cohen-Walker-Witteborn" framework of 
absolute photometry and stellar spectra is assured
(Cohen, Walker, \& Witteborn 1992a).  The method is 
based on the use of either ``supertemplates", that represent the intrinsic 
shapes of the spectra of K0--M0III stars from far-ultraviolet (1150\AA) 
to MIR (35~$\mu$m) wavelengths, or Kurucz synthetic spectra
for A0-5V stars.  Each supertemplate or Kurucz model is reddened according to the
individual star's extinction and is normalized using available characterized 
optical photometry.  This paper tests our capability to predict NIR ($JHK$) 
magnitudes using supertemplates or models constrained by Hipparcos/Tycho or
precision ground-based optical data.
We provide absolutely calibrated 0.275--35.00~$\mu$m spectra of thirty three Landolt or
Carter-Meadows optical standard stars to demonstrate the viability of this technique,
and to offer a set of IR calibrators 100-1000 times fainter than those we have
previously published.  As an indication of what we can expect for actual IRAC calibration
stars, we have calculated the absolute uncertainties associated with predicting the
IRAC magnitudes for the faintest cool giant and hot dwarf in this new set of calibration stars.
\end{abstract}


\section{Introduction}
This series of papers on spectral irradiance calibration in the infrared (IR) 
was motivated by the need to establish accurate celestial flux standards for 
use in astronomical spectroscopy and photometry by spaceborne, airborne and 
ground-based instruments.  The earlier papers in this series present the
foundations of a method for establishing an entire all-sky network of IR 
flux calibrators.  The NIR to MIR imager on NASA's Space InfraRed Telescope Facility 
(SIRTF) - the Infrared Array Camera (IRAC) - has four detector arrays, with 
central wavelengths of 3.6, 4.5, 5.8 and 8.0~$\mu$m.  The IRAC detectors are 
roughly 2000 times more sensitive than those of IRAS and would saturate on 
all but the faintest stars of the existing network.  In the current paper, we 
develop the methodology to predict the IR in-band fluxes of fainter stars 
using optical photometry, and we report the results of an experiment to 
demonstrate the reliability of this method.  Consequently, this paper 
describes the approach we are pursuing to create the on-orbit cool giant and 
hot dwarf calibrators for IRAC, and will serve as a proof-of-concept by 
validating the efficacy of the method.

The objectives of the current paper were to: 
select a mix of cool giants and A-dwarfs with well-characterized, precision 
optical photometry; determine accurate optical spectral types for all these 
stars; represent their energy distributions by appropriately reddened 
stellar photospheric spectra; normalize these energy distributions using the 
optical bands; predict $JHK$ from the normalized spectra; secure 
characterized $JHK$ observations of as many of these stars as possible;
compare observed and predicted NIR magnitudes; and create a set of absolute
spectra for stars considerably fainter than those of our previous all-sky
network.  The final products to be generated by this effort will likewise
be complete, continuous, absolutely calibrated 
spectra from the UV to MIR, consistent with the context described by Cohen et 
al. (1999).

\S2 overviews the basic components of our methodology, treating: the 
construction of combined optical--IR ``supertemplates" of cool, normal giant 
stars, which extend from 0.115 to 35~$\mu$m; the optical--IR atmospheric models 
and their synthetic spectra 
for the hot A-dwarfs; and the reddening corrections applied to the selected 
stars.  \S3 follows with a demonstration of the method using a set of 
selected Landolt and Carter-Meadows standards with published, precision $UBVRI$
magnitudes, for which we secured our own optical classification spectroscopy,
and applies the technique to the prediction of NIR and
MIR magnitudes from optical photometry and spectra. \S4  describes the  results
of renormalizing these supertemplate and model spectra
 using all available optical, NIR, and MIR photometry, and describes how to
obtain these spectra. \S5 sets our work in context with 2MASS and SIRTF, showing how this 
method will be used to estimate the in-band fluxes of our potential calibrators 
in the four IRAC bands and assessing the degree to which this approach will 
satisfy the absolute calibration goals for IRAC.  In
an Appendix, we detail the extension of the method to accommodate space-based optical data
from Hipparcos/Tycho, and we focus on the problems encountered with
Tycho-1 photometry of cool stars, which are largely mitigated by Tycho-2 data.

\section{Methodology}
\subsection{Overview}
The method extends previously created, empirical, absolutely calibrated, 
1.2--35~$\mu$m composite spectra (``templates") of bright stars with spectral 
types of K0III to M0III, into the UV and optical.  These optical-to-infrared 
``supertemplates" are used as generic spectral shapes for all stars of a given
spectral type and luminosity class.  The supertemplates, after they have been
reddened using a standard extinction law and normalized to broadband optical
and NIR/MIR photometric measurements, serve as proxy spectra for a network of 
stars which are too faint or too numerous to be effectively measured 
individually.  

\subsection{Limitations of the current network of stars}
The current network of 422 stars was
constructed by fitting 1.2--35~$\mu$m templates to photometry from
ground-based and spaceborne telescopes (Cohen et al. 1999).  The
fundamental reference standards for the spectra are absolutely calibrated
models of the A-type stars Sirius and Vega.  The templates were constructed by 
combining empirical spectral fragments for eight bright K/M-giants obtained 
from ground--based, airborne and spaceborne telescopes.  The absolute
calibration of these spectra was determined by taking the ratios of the
cool stars' spectra to the observed spectrum of either Vega (below 13$\mu$m) or Sirius, and
then scaling the ratioed spectra by the theoretical A-star spectra, degraded
to the actual spectral resolution of the observations.

The range of fluxes in the network of 422 templates decreases from 800 to 5 Jy in
the IRAS 12-$\mu$m band, with a handful of stars as faint as 1.1 Jy in a
new, unpublished set of 602 templates by Walker \& Cohen (2002) which includes
the 422 stars of Paper X, sometimes with updated versions of their spectra.  The
MSX satellite has recently provided a validation of the relative and absolute 
calibration of subsets of the stars that compose this network.  Cohen et al. 
(2001) have analyzed independently
calibrated MSX photometry in six bands from 4.3 to 21.3~$\mu$m from all
three tiers of the network - Vega/Sirius, the bright K-/M-giants, and
the fainter template stars - and find the MSX data agree with the
predicted fluxes to within the estimated uncertainties.

To provide meaningful calibrators for IRAC's four bands requires flux densities 
from about 1 Jy down to about 1 mJy in these bands.  However, we wish 
to maintain traceability to our absolute spectral products that 
support the Infrared Telescope in Space (Murakami et al. 1996), the 
Diffuse Infrared Background Explorer (Hauser et al. 1998, their Table~1; 
Cohen 1998), the Midcourse Space
Experiment (Price et al. 2001; Egan et al. 1999; Mill et al. 1994)
and ESA's Infrared Space Observatory (Kessler et al. 1996) instruments.  To
extend K0--M0IIIs from 12-$\mu$m flux densities of 5 Jy down to the
requisite values for IRAC necessitates using stars that are fainter by about 9 
mag than those of the current network, at the level of V$\sim$11--12.  One way 
to achieve even fainter IR flux densities, for stars with the same visual 
magnitudes as the current network, is to use hot stars, for which Kurucz model 
atmospheres are quite reliable in the optical and IR.  

\subsection{Creating new ``templates'' and ``supertemplates'' for K-/M-giants}
The original set of spectral types for which Cohen et al. (1995,1996a,b) have 
constructed complete, empirical 1.2--35~$\mu$m spectra consists of: K0, K1.5, K3,
K5, M0, M1.5, M2.5, and M3.4III.  To minimize any potential variability we 
have concentrated solely on the range K0-M0III for IRAC calibrators.  To 
create a finer grid of templates, we have interpolated these spectra to create 
new IR templates for types K1, K2, K4, and K7III.

The interpolation of a template file, with its five elements 
($\lambda$, F$_{\lambda}$, absolute error in F$_{\lambda}$, local, and
global biases: see Paper X) first requires removal of the 
global and local biases from the total uncertainty, leaving the solely 
random component of the uncertainty.  Two templates whose spectral types
flank that of the required type are normalized in $\lambda^4$F$_{\lambda}$ 
space so that they lie on top of one another longward of the SiO fundamental.  
The flux density, random error, local bias and global bias are linearly 
interpolated using the {\it s} parameter of de Jager \& Nieuwenhuijzen 
(1987), which represents spectral type more continously and linearly than 
either stellar effective temperature or its logarithm for normal, mature
stars.  Finally, the local 
and global biases are recombined in quadrature with the random errors,
yielding the quantities for a new template file.  This procedure provides 
a more complete set of template spectra to apply to new candidate calibrators,
and a set that it is fully consistent with
our published, complete, empirical spectra of bright K-/M-giants.  Figure~\ref{k4g}
illustrates the newly created K4III template derived by interpolation between K3III
($\alpha$ Hya) and K5III ($\alpha$ Tau) templates.

To establish the credibility of such interpolated products, we compare the 
K4III template with consistently calibrated spectral observations of $\beta$ 
UMi (K4III) 
taken on the Kuiper Airborne Observatory (KAO) using the HIFOGS instrument 
(Witteborn et al. 1995) and by the IRAS Low Resolution Spectrometer (LRS).
Both HIFOGS (Cohen et al. 1995) and the LRS (Cohen et al. 1992a) 
have been calibrated in the identical context to that of Paper X.
$\beta$ UMi was observed on the KAO flights of 14 and 19 April 1995, and
calibrated using identically taken HIFOGS spectra of $\alpha$ Boo, 
matching the airmass of $\beta$ UMi as closely as possible.
Figure~\ref{kaocfn} illustrates this direct comparison.  Note that the two
spectrometers have very different resolving powers, of order 200
for HIFOGS (4.9--9.6~$\mu$m) and 20-50 for the LRS (7.7--22.7~$\mu$m).
This accounts for the divergence, for 7.7--8.5~$\mu$m, between the filled 
(KAO) and open (LRS) squares in the deep SiO fundamental absorption.
With these exceptions, there is satisfactory overlap
at the 1-2$\sigma$ level between observations and the K4III template,
corresponding to agreement within 5\% (except in the saturated ozone
feature $\sim9.3~\mu$m in the KAO spectrum).

To construct combined optical-to-IR ``supertemplates", it is necessary 
to extend the standard 1.2--35~$\mu$m calibrated spectra into the optical.  
This step was accomplished by splicing the IR templates to
the average of the several spectra
for stars with relevant MK spectral types from Pickles's (1998) spectral 
library that extends from the far-ultraviolet (1150\AA) to the NIR
(2.5~$\mu$m).  In some cases, it was necessary to fill in ``gaps'' in the 
Pickles ``UVKLIB'' spectra of the cool giants and to rectify erroneous blackbody-based 
interpolations by using our own fully-observed
spectra longward of 1.22~$\mu$m.  When the final products were deemed  
complete, accurate, and without discontinuities, the observed 
$(B-V)$ colors for these supertemplates were synthesized and compared with
the literature to check that they satisfactorily represented reddening-free 
spectral shapes, within the respective uncertainties.  Figure~\ref{ste} presents the 
K5III supertemplate from 0.27--35~$\mu$m.  Supertemplates formally extend down 
to 0.115~$\mu$m but cool giants rarely are well-detected by IUE below 0.27~$\mu$m,
so we offer them only longward of 0.275~$\mu$m, for practical purposes.
Each Pickles spectrum was itself constructed as the average of spectra from
up to 17 other libraries, suggesting probable robustness as a representative
for each type.

Although we created a K7III template, we recognize that these are rather
rarely encountered and, further, Pickles (1998) does not include this
type in his spectral library.  Therefore, we simply averaged his spectra
for K5III and M0III, and appended that to our interpolated K7III IR template.

Implicit in these constructs is the assumption that whatever the actual
metallicity and, more importantly, the abundances of C, O, and Si are for any
individual star, we can apply a generic spectral shape whose heritage is
partly traceable to the bright stars observed spectroscopically from the KAO, 
from the ground, and from space (using the LRS).  The reason for this 
assumption is that this same technique will be applied to visually very faint 
stars for which it is most unlikely that such detailed information will be 
available.  

\subsection{Atmospheric models and synthetic spectra of the hot stars}
The A-dwarf synthetic spectra were all taken from the standard, solar-abundance grid of 
Kurucz LTE model atmospheres (Kurucz 1993), 
and correspond to stars of effective temperature, 9795, 9397, 9016, 8710,    
8433, 8185K, for the set A0V to A5V, respectively (de Jager \& Nieuwenhuijzen
1987), and gravity, log(g), between 4.16 and 4.25 (varying monotonically with type).

\subsection{Supertemplates: validations and usage}
Analysis of both empirical IR spectra and theoretical models indicates that effective
temperature exerts the dominant influence on the shapes of these spectra, with
gravity, and especially metallicity, much less significant.  Tests carried out on Kurucz
models (for hot, warm, and even cool stars) during the calibration work in support
of ISO showed that differences arose between IR templates and models, even
when the literature provided explicit estimates of temperature, gravity and [Fe/H].
However, empirical and model spectra were generally in good agreement, within $\pm$2\% 
in the IR continuum, and $<$5\% in molecular bands (e.g. the CO first overtone bands).
Further, those differences that do arise were not dependent on [Fe/H] but rather
on a star's individual abundances of C, O, and Si.  Such data are not
generally available for faint stars, even when an estimate of [Fe/H] might exist.

Another way of validating this entire process derives from the direct comparison of
observed stellar angular diameters and ``radiometric" diameters (Cohen et al. 1999).
There is excellent agreement between measured
and radiometrically predicted stellar angular diameters, with the implication that
continua and even molecular band shapes and depths are well-matched at least
to first order.  This result also lends validity to the supertemplates.

\subsection{Reddening corrections}
Stellar reddening was determined by comparing Landolt's (1992) measured 
$(B-V)$ (or, equivalently, the same indices measured by Carter \& Meadows
(1995)) with the mean intrinsic colors described below.  If a star proved too 
blue for its type, we assigned zero to A$_V$, otherwise we assigned 
3.10$E(B-V)$.  The resulting extinctions were applied to the intrinsic 
supertemplates.

\subsubsection{Intrinsic $(B-V)$ colors}
The natural spread of observed $(B-V)$ indices for unreddened stars of any 
given spectral type is quite substantial, attaining about $\pm$0.15
(Mermilliod 1993).  We attribute this to intrinsic cosmic scatter among the
stars, caused by: variations in metallicity; the individual abundances of
elements - primarily C, O and Si -  whose lines and bands significantly sculpt
the stellar energy distribution and are not governed solely by temperature
and gravity; the quantization of a continuum of stellar
temperatures and gravities into discrete spectral classes; and errors
in the assignment of spectral types drawn from the literature.  It is
for this latter reason that we have mounted our own optical spectroscopy
program (see \S3.4).

To exact the greatest precision from our technique requires the best
estimates for the mean color indices of K0--M0III and A0--5V stars
because optically faint (hence IR-faint) stars of both these types are
likely to be significantly reddened.  For these estimates we chose to rely 
on the Hipparcos output data base (van Leeuwen 1997). Individual records offer 
the Hipparcos team's best value of $(B-V)$ color, along with kinematic and 
parallax data, spectral types, and a host of information so that one can 
readily reject multiple and variable stars, and objects with poorly determined 
parallaxes.  The catalog enables the determination of observed colors from 
subsets with as many as 50--200 stars for each populous type of K-giant, and at 
least 50 A-dwarfs per subclass, even when demanding meaningful measurements of 
parallax (hence distance) at the 5--10$\sigma$ level.  

We converted observed $B$ and $V$ magnitudes to absolute $M_B$ and $M_V$ and
into estimates of each star's intrinsic $(B-V)$.  If a star's distance, 
calculated from the reciprocal of a significant parallax, placed it within
the dust-free local bubble (Fitzgerald 1970; Perry \& Johnston 1982; Perry 
et al. 1982) then we assigned zero extinction on the basis of the extent of the 
dust-free zone.  For stars beyond 75 pc, we assigned an A$_V$ of 0.625 
mag kpc$^{-1}$ (and A$_B$/A$_V$=1.299) to determine the absolute magnitudes and
intrinsic $(B-V)$.  The
values used for the extinction per kpc, and for the ratio A$_B$/A$_V$, come
from the reddening law described below.  
Because we test each parallax-based distance against a 
simplistic representation of the local bubble, then average the resulting
extinctions (many of which are zero), any residual effects of Lutz-Kelker bias
are greatly diluted.

\subsubsection{The reddening law}
The representation of the actual law of extinction that we used between FUV
and MIR was
based on the 7th-order polynomial fits suggested by Cardelli, Clayton \& Mathis 
(1989) for the ultraviolet and far-ultraviolet, the 8th-order polynomial fits of
O'Donnell (1994) for the optical-to-NIR range, and joins smoothly onto 
the law of reddening used by Cohen (1993) longward of 4.7~$\mu$m. 
We note that there are few substantive differences at short wavelengths 
($\leq1.0\mu$m) between different, empirically-derived, reddening laws in the 
literature.  This overall law compares favorably, for example, with the 
representations of other authors, such as Fluks et al. (1994), whose IR
reddening law is actually derived from theoretical considerations.
We have likewise compared our reddening curve
with that described by Fitzpatrick (1999), finding excellent accord except 
in the vicinity of the $R$ band (see Fitzpatrick's discussion of this region)
and, of course, throughout the interstellar silicate absorptions that are
absent from his representation of the IR extinction curve.  However,
Fitzpatrick's (1999) quantification of the uncertainties in UV-optical
extinction curves is particularly valuable.
To estimate uncertainties in the NIR and MIR, we first compared the law 
described above with the independently derived law of Rieke \& Lebofsky (1985). 
Following their adoption of uncertainties of $\sim$15\% in A$_{\lambda}$/A$_V$,
we find that the differences between our values and those of the 21 Rieke-Lebofsky 
points lie well within 1$\sigma$ (combined) for 18 wavelengths.  Only one point
(at 13~$\mu$m) lies more than 2.5$\sigma$ away.  
These quantitative uncertainties in our reddening curve are now
accommodated in the code that generated the actual spectra of the Landolt and
Carter-Meadows stars in this paper, and will likewise be used to create spectra
of the SIRTF calibrators.

\section{A demonstration and proof-of-concept}
\subsection{Overview}
We now offer a demonstration of our methodology using a set of 
selected optical standard stars with published, precision $UBVRI$ magnitudes,
and for which we have undertaken an optical spectroscopy program at Mt.  Hopkins 
Observatory (MHO) to establish their spectral types.  Mt.  Hopkins spectral types and
extinctions based on the observed $(B-V)$ are used to appropriately redden
the supertemplates or Kurucz synthetic spectra to represent each selected 
star's energy distribution.  These shapes are then normalized by $UBVRI$
photometry and the resulting spectra used to predict IR $JHK$ magnitudes.
Direct comparison between well-characterized NIR photometry and these
predictions tests the efficacy of our approach.  

\subsection{Selection of stars} 
We now describe our chosen stars which were drawn from two different sets, 
with their own independent optical and NIR photometry, and discuss the sources 
of their spectral types.  We offer cross-checks on two independent $UBVRI$ 
data sets; 
compare space-based and ground-based optical normalizations of supertemplates; 
and demonstrate consistent predictions of different systems of $JHK$ for one 
star common to the two data sets.

Our primary optical data set is the precision $UBVRI$ photometry on 
stars in the Kapteyn Selected Areas offered by Landolt (1973,1983,1992).
These furnish an abundant set of faint, optical standards measured through 
well-documented passbands.  Only about 600 of the many stars observed by
Landolt had ever been classified spectroscopically, by Drilling \& Landolt 
(1979).  Of these, only about 30 were suggested to be K0--M0 giants.  Early 
experiments with the technique described in this paper indicated that some of 
these types were surely suspect by several subclasses.  Consequently, we 
selected 32 alleged cool giants and about an equal number of stars
lying within 10$^\circ$ of the Ecliptic Plane, whose $(B-V)$ colors suggested 
that they might be K0-M0IIIs, for a program of optical classification
spectroscopy.  In total we investigated 62 Landolt stars.

Another valuable set of test objects is the collection of stars offered by
Carter \& Meadows (1995) as faint standards for 3--4m class ground-based
telescopes.  Carter \& Meadows (1995) also provide their own $UBVRI$ measurements 
from the SAAO 0.5-m telescope at Sutherland.  To test the A-stars, we decided 
to apply our method to {\it all} stars for which Carter \& Meadows cite a 
spectral type between ``A0" and ``A5", even when a luminosity class is lacking.  
We had hoped to demonstrate the plausibility of assigning dwarf luminosities 
to these stars on the basis of a successful prediction of their NIR magnitudes.  
This gave us an additional 19 stars to investigate.

We have augmented the Landolt sample of cool giants for which we 
obtained Mt. Hopkins spectra by SA94-251, SA108-475, 
and HD197806 (whose type of K0III, cited by Carter \& Meadows, 
is vindicated by Houk \& Swift (1999)).

We found a final subset of 24 cool giants stars to be useful: i.e. their 
actual spectral types fell into our desired spectral ranges; we were able to 
secure precision NIR photometry from Tenerife or from Carter \& Meadows; and 
we could accurately predict their $JHK$ magnitudes.  These 24 stars appear in 
Table~\ref{coords}.  The other Landolt stars were ``lost" either because we 
found them to be K- or M-dwarfs, or because we were unable to secure 
Tenerife $JHK$ data for them.

\begin{table}
\dummytable\label{coords}
\end{table}

We extracted the most accurate coordinates that could be found either in 
SIMBAD (``S") or in the Guide Star Catalog GSC1.3 with corrections from ACT 
(``G") for these stars.  Searching SIMBAD by any of a host of possible star 
names is not equivalent to searching by coordinates, as Dr. David Shupe, who 
kindly extracted these coordinates for our first observing run, noted for
SA114-670.  The SIMBAD position returned by this name is considerably in error 
compared with the true coordinates that correctly return the star under
the name PPM 700910.  
Table~\ref{coords} summarizes accurate coordinates, with Landolt or 
Carter-Meadows $B$ and $V$ magnitudes.  Carter \& Meadows provide rough J2000 
positions of their stars; but their A-dwarfs are all HD stars and precise coordinates 
are available unambiguously in the literature.

\subsection{Optical photometric bands}
Dr. Arlo Landolt kindly provided digitized versions of the tables from his 
1992 paper in which are given details of the CTIO $UBVRI$ passbands used for 
his photometry of optical standard stars, along with the response curve of the 
photomultipliers.  We further multiplied the product of each passband and
the detector response curve by a mean atmospheric transmission spectrum
appropriate to Cerro Tololo Inter-American Observatory (CTIO) in Chile.  To represent 
the intervening telluric transmissions above CTIO, we represented the atmospheric
transmission using PLEXUS\footnote{http://www2.bc.edu/~sullivab/soft/plexus.html\#Desc}, 
an AFRL validated ``expert system" that incorporates atmospheric code, specifically 
MODTRAN 3.7, SAMM, and FASCODE3P with the HITRAN98 archives.
PLEXUS contains an extensive database to support its expert aspect, so that
the effects of Rayleigh scattering, aerosols, and particulates appropriate to the 
desert conditions of the CTIO sites were included.  Paper X also used PLEXUS calculations 
to represent the site-specific atmospheric transmissions necessary to represent
the tens of ground-based filters characterized in that work so we have maintained
consistency in our treatment of all systems, both optical and IR.  We did compare
the CTIO atmospheric transmission from PLEXUS with the simpler formulation for extinction 
given by Hayes \& Latham (1975), based explicitly on the effects of Rayleigh
scattering, absorption by ozone, and scattering by aerosols, and we found good accord.

We created $UBVRI$ relative spectral response curves (hereafter RSRs) from the
combinations of filter, atmosphere, and detector profiles that were normalized 
to peak values of unity.  There have been several
comparisons of the SAAO $UBVRI$ system with Johnson's $UBVRI$ (Cousins 1984;
Bessell 1979), CTIO $VRI$ (Menzies 1989), and a wider ranging discussion is
given by Straizys (1992).  Menzies (1989) found slight differences between the Cousins 
$RI$ passbands and those of Landolt, while Menzies et al. (1991) concluded that
the SAAO and Landolt CTIO $UBV$ bands differ from one another and probably also
from Johnson's.  We have attempted to pursue the actual bands in use at SAAO by
way of the detailed response functions offered by Straizys (1992: who also cites
Bessell 1979,1983), but this has proved unsatisfactory.  We further note the following 
statement by Menzies (1989):
``The photometer on the 0.5-m telescope in Sutherland is in continuous use so
there has not been an opportunity to determine the filter transmission functions
experimentally."  

Consequently, we have represented the Carter-Meadows relative response curves 
(filter-plus-atmosphere-plus-detector) in $UBVRI$ by the same response curves as those 
published by Landolt (1992).  Direct comparison of $UBVRI$ for objects in common to
the Carter-Meadows and Landolt samples justifies this step, as exemplified
by SA94-251 and SA108-475 (Table~\ref{cfns}).  For the former, the differences between
magnitudes measured in both systems are within the assigned 3$\sigma$ uncertainties;
for the latter, all differences are within the 1$\sigma$ uncertainties.

Menzies et al. (1991) based their conclusions on a set of 212 Landolt stars for which
they present SAAO $UBVRI$ measurements.  These stars span a wide range of spectral types.
We have carried out the same experiment, but restricted to 21 Landolt stars in common, for which
our Mt.  Hopkins spectra indicate K0-M0III spectral types.  The result of comparing
Landolt-minus-SAAO magnitudes, using inverse-variance weighting for the combinations,
show, for $UBVRI$ respectively: -0.002$\pm$0.022; -0.012$\pm$0.011; -0.005$\pm$0.007;
-0.011$\pm$0.009; and -0.011$\pm$0.010.  On the basis of this restricted range of
spectral types relevant to our needs, we conclude that there are no significant 
differences between Landolt and SAAO photometry.  This conclusion also enables us
to utilize the data in Table 1 of Menzies et al. (1991) with our subset of Landolt stars.
For simplicity, when we later refer to ``Landolt'' of ``$BVRI$ photometry", this 
might include both Landolt's and Menzies' measurements.

\begin{table}
\dummytable\label{cfns}
\end{table}

All five system response curves were then integrated over our standard, 
calibrated, Kurucz model spectrum of Vega (Cohen et al. 1992b) to provide
their ``zero magnitude attributes".  These attributes (Table~\ref{gndcal})
include the in-band flux (irradiance) and its uncertainty, and the
monochromatic specific intensity (e.g. F$_{\lambda}$), as well as the
isophotal wavelength for Vega of each filter.   For convenience we have also
incorporated F$_{\nu}$, the monochromatic specific intensity in frequency
terms, in units of Janskys (Jy).  Note that the isophotal F$_{\nu}$ cannot be accurately 
rendered simply as 
$\lambda$$_{iso}$[$\mu$m]$^2$~$\times$~F$_{\lambda}$[W cm$^{-2}$ $\mu$m$^{-1}$]/3$\times$10$^{-16}$ 
because of the finite bandwidths of these filters.  We have recast each RSR in frequency terms,
and have performed the identical integrals for in-band, bandwidth, and F$_{\nu}$ 
as we carried out for F$_{\lambda}$.  The electronic version of 
Table~\ref{gndcal} offers these combined RSRs by clicking on the links
in the ``Filter" column.  

While Cohen and 
colleagues (e.g. Paper X) have always defined zero magnitude in the
infrared by this Kurucz spectrum of Vega, this star does
not have zero magnitudes in the optical.  We adopted magnitudes 
for Vega of $U$=0.024, $B$=0.028, $V$=0.030, $R$=0.038, $I$=0.034
(Bessell et al. 1998).  Therefore, we derived the true zero magnitude
irradiance and isophotal F$_{\lambda}$ values from those of our
Vega spectrum, brightening the Vega values accordingly.
Table~\ref{gndcal} lists the resulting zero magnitude 
attributes for the five bands consistent with our 
previously published absolute calibrations in the optical and IR. 
Landolt's system of photometry is based on Johnson's
magnitudes and consequently corresponds to the above magnitudes for
Vega.  Therefore, we have applied no zero point offsets to align his 
photometry with our definition of the zero magnitude attributes.

The final column of Table~\ref{gndcal} gives the monochromatic AB$_{\nu}$
magnitudes, as defined by Oke \& Gunn (1983), that correspond to zero magnitude 
for each of the five bands (sometimes described as ``zero point magnitudes").

\begin{table}
\dummytable\label{gndcal}
\end{table}

\subsection{Our classification spectroscopic program}
Our 62 selected Landolt stars 
were observed with the FAST spectometer on the 1.5 meter Tillinghast telescope of the Fred
Lawrence Whipple Observatory (Fabricant et al. 1998).  The spectrometer was
configured to provide 4000--7000\AA\ spectra with 5\AA\ resolution.
The stars were observed as part of a queue observing program on eight
different nights, under conditions ranging from photometric to thin
clouds, at airmasses below 1.64.  Exposure times ranged from 0.5 to
1800 seconds. A $2^{\prime\prime}$ slit was used to obtain 5\AA\ resolution. 
The data were dark-subtracted, flat-fielded, and one-dimensional spectra were
extracted.  These spectra were then subjected to background subtraction using the
adjacent off-source spectra in the original two-dimensional spectra, and a spike
rejection routine was applied to eliminate cosmic ray hits.  The
wavelength scale was determined using calibration lamps.

To establish a methodology for spectral types that was consistent with the MK
classification system, spectra of 34 MK standards (24 cool giants, 8 cool dwarfs 
and 2 cool supergiants, drawn from the Perkins Revised types of Keenan \& McNeil
(1989)) were obtained in
parallel with the targeted 62 Landolt objects, using
an identical instrument set-up and data reduction procedure.  These
spectra, displayed in Figures~\ref{mho1} and \ref{mho2}, show that the FAST spectra, when ordered
by spectral type and luminosity class, exhibit a systematic progression
of spectral features.  This demonstrates that accurate spectral
classification consistent with the MK system can be obtained with FAST
spectra.  Based on independent types in the two wavelength regions,
$\lambda\lambda$4000-5500\AA\ and $\lambda\lambda$5500-7000\AA\ at our $\sim$5\AA\
resolution, we estimate the uncertainty in our assigned types to be better than 
$\pm$1 spectral subclass.  

The release of Volume 5 of the Michigan Catalogue of Two-dimensional Spectral 
Types for the HD Stars (Houk \& Swift 1999) extends 
the previous four southern hemisphere volumes across the celestial
equator.  This systematic reclassification in the MK system has enabled us 
to compare our Mt. Hopkins types for 14 of the brighter Landolt stars that 
have HD numbers, with the Michigan types.  For these K- and M-giants, our 
types do not differ from the Michigan types by more than 1 subclass, 
validating our estimated uncertainty in spectral type. Furthermore, in classifying 
the Landolt cool giants we have become particularly aware of the relevance of 
the Michigan philosophy whereby one assigns, for example, a type of ``K2/3III" 
as opposed to K2.5III, with the implication that the star in question does not 
resemble in every way an MK-type of K2.5III, but rather that different 
criteria are found in the spectrum, some characteristic of a K2III and others 
of a K3III.

Volume 5 of the Michigan Catalogue now offers two-dimensional types for many of 
the Carter-Meadows A-stars, a number of which are indeed assigned to dwarf 
luminosity class, the bulk of them indicating agreement (to within 1 subclass)
with the types Carter \& Meadows (1995) drew from the literature.

\subsection{Predicting NIR flux densities}
\subsubsection{New NIR photometry of faint standards}
To test the efficacy of predicting NIR magnitudes from these optical data, we 
secured $JHK$ photometry using the 1.5-m 
TCS\footnote{The TCS is operated on the island of Tenerife by the Instituto de
Astrof\'{\i}sica de Canarias at the Spanish Observatorio del Teide of
the Instituto de Astrof\'{\i}sica de Canarias.} at Izana, Tenerife.  Paper X (in 
its Table 2 under ``Tenerife", and Table 3 under ``P.H.") 
presents the zero point offsets and absolute
attributes of this well-characterized set of filters. Table~\ref{tcs} presents
our new NIR photometry for the Landolt stars that we have so far been able 
to measure from Tenerife, together with their uncertainties.  All but one star 
were measured on at least two nights in at least one observing period (either 1998 May 
23-24 or 1999 July 12-14), and we obtained good accord between data from the 
two separate observing runs (Table~\ref{tcs}).  Data for stars observed at two epochs
have been combined using inverse-variance weighting, and the combined data 
also appear in Table~\ref{tcs}. The alternative route depends on Carter-Meadows $JHK$
data from SAAO, an equally well-characterized photometric system, also detailed by
Paper X (in its Table 2 under ``SAAO", and Table 3 under ``B.C.").

\begin{table}
\dummytable\label{tcs}
\end{table}

For each hot stellar model or cool stellar supertemplate deemed relevant to a 
particular star, we assembled all available precision optical photometry, whether
due to Landolt, Carter-Meadows, or Menzies.  We then integrated all five $UBVRI$ response curves over 
the reddened spectrum (or the two (Tycho-only) or three (Tycho and Hipparcos) 
spaceborne optical response curves likewise: see Appendix).  The resulting 
integrals through each system band provide the in-band flux and, converting 
the observed input magnitude into an equivalent irradiance through Table~\ref{gndcal}, 
we derive a scale factor {\it for each filter} in order to match 
the observed and normalized irradiance values.  Each overall set of optical 
data (five or ten ground-based, and/or two or three space-based magnitudes) yields an 
inverse-variance-weighted mean scale factor for the reddened supertemplate 
(using the photometric uncertainties that constitute an essential element of 
our approach), and a fractional uncertainty in this mean multiplier, termed 
the ``supertemplate bias", and expressed as a percentage of the mean scale 
factor.  We limit the magnitude uncertainties to be $\geq$0.005, 
for all optical and NIR observations, to avoid any single filter overwhelming 
the weighted scale factor.  Once the supertemplate bias is available, this
quantity is combined in quadrature with the global bias in the original
supertemplate shape, and thus finds its way into the total wavelength-dependent
errors for any star.

Table~\ref{112275opt} illustrates the process of determining the scale factor for a
reddened supertemplate for the star, SA112-275, a K0III with A$_V$=0.620. Column (1)
indicates the filter used; col.(2) gives the scale factor to match predicted to observed
in-band fluxes; col.(3) shows the fractional uncertainty in that scale factor; col.(4)
provides the isophotal wavelength associated with the combination of filter and
supertemplate.  The final line summarizes the inverse-variance weighted mean scale 
for this star based on the five individual scale factors.  Note that this table is
based solely on Landolt $UBVRI$, whereas the final version of the optically
normalized supertemplate for this star (and for many others in this paper) will be 
constrained using photometry from Landolt, Menzies, Hipparcos, and Tycho.

As Table~\ref{112275opt} shows, the scale factor derived from the U-band data is 
obviously much smaller than those associated with the $BVRI$ bands.  On detailed
examination, we found this situation occurred for almost every star.  Possible
explanations for this kind of behavior are discussed by Bessell (1990; his \S6).
Therefore, because we
wished these stars to be treated in the most precise way possible, we decided not to
use the $U$ photometry for any of the sample of stars.  Utilizing solely $BVRI$, but
combining both Landolt's and Menzies' data, we obtained a scale factor of
5.628E-04$\pm$1.860E-06, 1.3\% larger than that in Table~\ref{112275opt}.  This
difference is typical of the rest of our sample of stars.

\begin{table}
\dummytable\label{112275opt}
\end{table}

\subsubsection{The prediction of NIR magnitudes}
To predict $JHK$, we integrated the Tenerife (TCS) NIR system response curves over 
the correctly-scaled and reddened supertemplates, and converted the NIR in-band 
fluxes into their corresponding TCS magnitudes.  For assessment of the accuracy of our
predictions we used the mean algebraic deviation (MAD) over the set of three TCS filters,
in the sense observed-minus-predicted $JHK$.  Ideally we would like to have the
NIR photometry points ``straddle" the normalized, reddened supertemplate,
rather than to achieve the smallest mean {\it absolute} deviation for the three points because 
that could arise by having a supertemplate lie entirely above, or below, the JHK data,
possibly implying an unsatisfactory bias.  Accurate prediction of $JHK$ from 
a $BVRI$-constrained supertemplate/model depends upon reliable estimates of spectral type,
extinction, supertemplate/model shapes, and both optical and NIR photometry, making
this a highly demanding process.

Table~\ref{112275tcs} shows the comparison of predicted and observed TCS $JHK$ magnitudes
from the final optically-constrained supertemplate for SA112-275 (i.e. based on data from
Landolt, Menzies, Hipparcos, and Tycho: see the Appendix for a discussion of how the Tycho 
photometry was handled), from which the calculation of MAD is performed. 
Column (1) shows the relevant
TCS filter; cols.(2) and (3) give the predicted magnitude and uncertainty after integrating
the RSR for that filter over the supertemplate; cols.(4) and (5)
offer the corresponding observed magnitude and uncertainty; and col.(6) lists the difference,
col.(4) minus col.(2).  The last line summarizes the MAD for the star 
as the unweighted mean of the numbers in col.(6). 

\begin{table}
\dummytable\label{112275tcs}
\end{table}

Table~\ref{stes} summarizes these details for all 24 K-/M-giants with their supertemplate types,
extinctions, final scale factors, biases, MADs, incorporating both Landolt/TCS and
Carter-Meadows/SAAO methods.  For SA108-475, both routes were used for our MHO 
spectral class of K3III; i.e., using Landolt $BVRI$ to predict Tenerife $JHK$, 
and using Carter-Meadows $BVRI$ to predict Carter-Meadows SAAO $JHK$.
While the two derived extinctions are not identical,
they are very close (0.028 mag apart), which may afford an independent method 
to assess the uncertainties in our derived values of A$_V$. 
We derived very similar scale factors and MADs, each pair differing only at the 
1.6~$\sigma$ level of the joint uncertainties.  Consequently, the third record
for this star represents the result of combining {\it all} its optical and NIR
photometry, and using an A$_V$ corresponding to the average of the two $(B-V)$
values.

The fact that the cool
giants have biases (the percentage uncertainty in the mean scale factor for
a supertemplate or model) that are all so homogeneous and well below 1\% 
reflects primarily the precision achieved in both the Landolt $BVRI$ and TCS $JHK$ data
sets, and secondarily the significant number of photometry points per star, between
8 and 16 (many stars have data from both Landolt and Menzies, and some also have 
Hipparcos/Tycho-2 data). The larger values seen for stars constrained by SAAO
photometry are essentially due to the more conservative uncertainties assigned
by Carter to his $JHK$ magnitudes ($\pm$0.025) as compared with those from Tenerife
(typically a factor of 2-3 times smaller).  With much sparser data sets, for 
example just $BVJHK$, typical of some potential IRAC calibrators, supertemplate 
biases are larger, typically $\sim$2-3\%.  

\begin{table}
\dummytable\label{stes}
\end{table}

Table~\ref{stesavs} similarly presents our results on nine A-dwarfs for
which the types cited by Carter \& Meadows are confirmed by the Michigan
Catalog.  The biases are all larger than those for the sample of cool giants
because we assign an uncertainty of $\pm$5\% to the synthetic spectra of
A-dwarfs, to account for the influence on the continuum and lines of
deviations in line strengths, gravities, metallicities, and real stellar
structures as compared with opacities and structures represented in models.
Examination of the MADs in Table~\ref{stes} and Table~\ref{stesavs} suggests the 
viability of predicting $JHK$ from optically normalized, reddened supertemplates
and models, chosen in accord with our MHO spectral types for cool giants, and 
using types drawn from the Michigan Catalogue for A-dwarfs.  Indeed, the 
(unweighted) average of the MADs for the ensemble of 33 stars is --0.007$\pm$0.005.  

\begin{table}
\dummytable\label{stesavs}
\end{table}

Figure~\ref{nirmads} presents the set of MADs for the A-dwarfs and cool giants
separately, broken down into the behavior at $J$, $H$, and $K$.  Every individual
star is plotted in the appropriate diagrams with the 1$\sigma$ error bars on the
values of MAD.  By and large, the sets of MADs are consistent with the ensembles
of stars having essentially zero offset between observation and prediction
at the 3$\sigma$ level, except perhaps for one or two stars.

\subsection{Predicting MIR flux densities}
\subsubsection{IRAS}
For nine Landolt stars we were also able to find IRAS measurements, either
from the Point Source Catalog (PSC) or, more often, the Faint Source Catalog (FSC) or 
Faint Source Reject Catalog (FSR).  These appear in Table~\ref{iras}, expressed as F$_\nu$ 
(in mJy).  For these stars, using the same scale factors as implied by the
optical normalizations, we predicted the 12- and 25-$\mu$m flux densities using the 
IRAS system response curves, and these flux densities are also given in Table~\ref{iras}.
In all cases, predictions lie within $\sim1\sigma$ of IRAS observations, 
further vindicating the efficacy of the technique used to create normalized 
supertemplates, and limiting the possibility that these cool giants might have 
circumstellar dust shells.

None of the Carter-Meadows A-dwarfs was detected by IRAS indicating that, at a rather
gross level, none of these stars is afflicted by the Vega phenomenon, in keeping
with their intended usage as calibrators.

\begin{table}
\dummytable\label{iras}
\end{table}

\subsubsection{MSX}
One star among the Landolt sample, SA114-176, was observed and
detected by MSX at 8.28~$\mu$m.  Table~\ref{msx} summarizes its observed
and predicted irradiances, isophotal flux densities, and magnitudes
in this MSX band, based on the scale factor for its
supertemplate determined from optical normalization.  All lie within 
$\sim$1.9$\sigma$ of the observed values, again validating our method.  
The final line in Table~\ref{msx} shows the predicted MSX quantities
based on normalizing the supertemplate for this star using all optical
and NIR data.  There is almost no difference between the predictions
determined from optical data, and from optical-plus-NIR data, essentially 
because the small MAD for this star (Table~\ref{stes}) indicates that the
two mean scale factors must be very similar.

\begin{table}
\dummytable\label{msx}
\end{table}

\section{The supertemplate files, and how to obtain them}
Having demonstrated the NIR predictability of these Landolt and Carter-Meadows stars
from the spectral types, supertemplates or models, and optical photometry, we have
renormalized each spectrum by utilizing every available optical, NIR and MIR photometric point 
to define the mean scale factor and bias for each templated star. 
These final spectra elevate our subset of the precision optical standards into IR calibrators,
and are self-consistent with our published set of almost 450 fiducial IR standards.
As an illustration of how the spectra are constrained by the photometry, Figure~\ref{montage}
shows supertemplates or models for 8 of the 33 stars.

The 33 supertemplates/models are available through the electronic
version of this paper. A detailed header precedes every supertemplate and model created 
as we have described.  This header is
accompanied by: various names for each star; release version, date, and time of spectrum
creation; the supertemplate spectrum used; the extinctions
of both the bright star that gave rise to the supertemplate shape and of
the templated star; for cool giants, the angular diameter deduced for the stellar
supertemplate (derived from the template scale factor and our published values
of angular diameters for the bright composite spectra: Cohen et al. (1996b));  
all the characterized photometry used to normalize the supertemplate,  
with references; and our determinations, for each filter used, of isophotal flux and its uncertainty,
and isophotal wavelength for the particular spectrum in question.  Supertemplates are always named for their HD
designations, when a star has an HD number, e.g. ``HD139513.tem" (for the star SA107-347); 
otherwise, the Selected Area star name will represent the file, e.g. ``SA113\_259.tem".

The actual calibrated stellar spectra have a five--column format (exactly as in Paper X). 
For the wavelength range from 0.275--35.00~$\mu$m,
we tabulate: wavelength ($\mu $m); monochromatic irradiance (F$_{\lambda}$ in units of
W cm$^{-2}$ $\mu $m$^{-1}$); total uncertainty (also in units of
W cm$^{-2}$ $\mu $m$^{-1}$) associated with this value of F$_{\lambda}$; local
bias; and global bias. For most applications, ``total uncertainty" is the error
term most appropriate to use.  It is the standard deviation of the spectral
irradiance and incorporates the local and global biases.  Local and global biases  
are given as percentages of the irradiance. The global bias does not contribute
error to flux ratios or color measurements, and may be removed (in quadrature)
from the total error.  Note that we prefer to provide pristine data, rather than to
regrid each supertemplate to an equally-spaced or common wavelength
scale.  Each supertemplate has a different set of wavelengths in the IR.  Consequently,
spectra are {\bf not} tabulated at equal intervals of wavelength but follow Pickles (1998)
as far as 2.500~$\mu$m, and then mimic the wavelengths of the originally observed 
composite spectra.

In the electronic version of this paper, the star names in the first column of 
Table~\ref{stes} are links to the supertemplates for the 24 Landolt stars (only
the last entry for SA108-475 has such a link), while Table~\ref{stesavs} similarly
links to the supertemplates for the 9 Carter-Meadows A-dwarfs.

Should other well-characterized optical or IR observations of any of these 33 
stars become available at some time in the future, that are not directly
calibrated by an earlier version of the same star's spectrum, it would be 
possible to recreate that supertemplate or model, hopefully with reduced
uncertainties.  This procedure was followed when MSX measurements first became 
available for a number of K-/M-giants from Paper X, resulting in the issuance
of updated spectra, with reduced template biases (Walker \& Cohen 2002).  The 
headers of the associated stellar spectra always carry the date of creation of 
the updated products so there will be confusion as to which generation of
calibration spectrum is represented when these files are downloaded.

\section{Conclusions}
The purpose of this paper has been threefold: to demonstrate the efficacy of our
approach to SIRTF/IRAC calibrators; to validate the resulting scale factors, spectral types, 
and extinctions using different optical data sets for a new, faint set of IR
calibrators; and to provide an archive of absolutely calibrated optical-to-infrared 
supertemplates/models for stars between two and three orders of magnitude fainter than
our current published network of IR standards.  We have been able to derive 
mean scale factors for a set of cool giant
supertemplates and hot dwarf models, drawn from a set of much fainter potential
calibration stars than we have previously established, with supertemplate/model
normalizations based solely on either ground-based or spaceborne optical
photometry.  We have demonstrated the capability of closely predicting the
NIR (and in a some cases even MIR) brightnesses of these stars using 
appropriately reddened supertemplates/models based on a modern set of MK spectral
classifications for these potential calibrators.  When applied to the actual calibrators 
for SIRTF, the combination of NIR and MIR points helps to reduce both random and systematic 
uncertainties in the 3-10~$\mu$m range covered by the IRAC.

For the present, we have offered a method that can successfully predict NIR $JHK$ magnitudes
for a sample of stars of well-determined spectral type in the ranges K0-M0III or A0-5V.  
Therefore, combining optical and NIR photometry (with MIR when available) on such stars
and renormalizing their supertemplates should satisfy our goal of having a set of absolutely 
calibrated optical-IR spectral energy distributions that will meet the uncertainties required
for IRAC calibrators.  We have attempted to quantify the uncertainties we expect for
potential calibrators by predicting the four IRAC magnitudes, and isophotal
flux densities (F$_{\lambda}$,F$_{\nu}$), and their uncertainties,
for the faintest A-dwarf (HD15911, A0V, K$\approx$9.5) and K-giant (SA114-656, K1III,
K$\approx$10.2) in our sample of thirty three stars.  Table~\ref{goals} presents this
information by band for each star.  The final column in the table offers the absolute
error in the in-band flux expressed in terms of the percentage fractional uncertainty.
These absolute errors include the estimated uncertainties in the IRAC relative spectral
response curves, in our supertemplates and Kurucz models, in the extinction law used,
and in the normalizing photometry.  All errors are well within the total error budget
of 10\% absolute set for IRAC.  For the cool giant stars,
molecular absorption bands contribute to IRAC2 and IRAC3 (the CO fundamental), and
IRAC4 (the SiO fundamental), driving up the empirical errors in these features in
the observations of the original composites from which supertemplates were derived.
We estimate that significantly fainter stars with poor and sparse photometry could have 
uncertainties of order $\sim$5\% in the IRAC bands.  However, even these errors leave a 
significant component of the budget to instrumental phenomena, such as the 
characterization of any non-linearities, and time-dependent responsivities on-orbit.

\begin{table}
\dummytable\label{goals}
\end{table}

In reality, although this work was undertaken to provide direct calibrational support 
for IRAC, there are additional constraints on the eventual SIRTF calibrators, namely that 
they should lie in the constant viewing zones surrounding each Ecliptic Pole.
None of the Selected Areas lies near the Ecliptic Poles.  However, an additional
solution we have proposed to meet IRAC's calibration needs is a subset of 
stars distributed in longitude around the Ecliptic Plane, within 10$^\circ$ 
of this plane in ecliptic latitude.  This extra set of standards would support calibration
either immediately prior to, or following, data downlinks, when SIRTF will lie in,
or close to, the Ecliptic Plane.  Some Selected Area stars, of relevand spectral
type and with Landolt photometry, lie within this Ecliptic Plane
zone.  Details of the Ecliptic Pole and Plane stars actually selected to support 
IRAC, and the derivation of their absolute spectra, will appear in a later paper in this series.

From the supertemplates, one can derive the expected attributes of the stars within any user-specified
set of passbands accommodated by the range of our spectra, 
given the system's relative spectral response curves, and one can utilize
the absolute spectra directly to support the calibration of low-resolution
(resolving power $\sim$100) spectrometers, exactly as for the 422 absolute template
spectra published by Cohen et al. (1999).  Thus, the supertemplate technique
offers the capability to predict IRAC magnitudes and to support low-resolution operations with 
SIRTF's InfraRed Spectrometer (IRS), thereby providing potential ``cross-calibrators" for 
this pair of instruments.

The acquisition of ground-based characterized $JHK$ photometry can be laborious,
particularly when one deals with potential calibration stars significantly fainter than those
discussed in the present paper, and some of SIRTF's calibrators lie in the southern hemisphere.
Therefore, we have made further use of these thirty three stars of intermediate
brightness, whose optical spectra are proven to be 
good predictors of NIR magnitudes, to extend our absolute
calibration framework to embrace 2MASS's $JHK_{s}$ bands (Cohen et al. 2003).
Thereafter, 2MASS photometry will provide an important resource for the assembly of even 
fainter calibrators for SIRTF.

\acknowledgments
MC's work on IRAC's calibrators is supported by contract SV9-69008 between 
UC Berkeley and The Smithsonian Astrophysical Observatory.  We are grateful
to Arlo Landolt for supplying electronically readable tables of his network of stars,
and of his filters and phototube response curves, and to Andrew Pickles
for his extremely helpful and detailed review.
We thank David Shupe for extracting the most accurate coordinates from SIMBAD for
the Landolt stars for our first observing runs at Mt. Hopkins, and
Mike Calkins and Perry Berlind for their invaluable efforts to secure FAST spectra.
We thank Fred Witteborn, Jesse Bregman, and Diane Wooden for the use of the 
unpublished HIFOGS KAO spectra of $\beta$ UMi from one of our collaborative flight
series in 1995.  This research has made use of the SIMBAD database, operated at CDS, 
Strasbourg, France.

\newpage	

\appendix
\section{Space-based visible photometry}
As a further test of the supertemplate method, and because we may well encounter
potential IRAC calibration stars that lack any characterized $UBV$ photometry,
we have compared independent 
normalizations of supertemplates solely from space-based optical photometry 
(Tycho $B_T$, $V_T$ [H{\o}g et al. 1997], and $H_p$ [from Hipparcos, when available])
against benchmark normalizations using solely Landolt's $BVRI$ data for 
stars in the celestial equator (to which the published Landolt
network is confined).
Bessell (2000) has discussed the problems and inadequacy of the published information that
accompanies the system response curves of the two Tycho bands ($B_T$,$V_T$) and the 
single Hipparcos band ($H_p$) given in Volume I of the Hipparcos archival 
volumes (ESA 1997).  We adopted Bessell's (2000) recommended relative spectral
response curves, which permit meaningful synthetic photometry.   Following
Bessell et al. (1998) and by interpolation, we adopted space-based optical magnitudes
for Vega of $B_T$=0.028, $V_T$=0.030, $H_p$=0.029. Table~\ref{htcal}
contains the zero magnitude attributes for the Hipparcos and Tycho bands, in the
identical manner as for the $UBVRI$ calibrations in Table~\ref{gndcal}. 

\begin{table}
\dummytable\label{htcal}
\end{table}

We compared mean scale factors (applied to supertemplates to match
predicted and observed irradiances) and their absolute uncertainties for cool giants,
for 12 Landolt stars and one Carter-Meadows giant, represented by supertemplates that are 
normalized by ground-based and by space-based optical photometry.
The scale factors derived for these giants from the space-based optical data are systematically
a few percent smaller than those from ground-based $BVRI$.  To investigate this phenomenon
in greater depth we have examined the individual scale factors derived from the $B_T$, $V_T$,
and $H_p$ bands.  Invariably, the $B_T$-scale is less than the $V_T$-scale, while the
$V_T$-scale is essentially equal to that found from $H_p$.  We note that this phenomenon is
not seen for the Carter-Meadows A-dwarfs for which the Hipparcos-Tycho bands 
available yield practically identical scale factors.  Further analysis reveals the fact that,
even for the cool giants, only the factors derived from $B_T$ are in question.  We note that
$B_T$ is strongly concentrated to the region around 4000\AA\ where cool giant spectra are
falling very steeply, below the Ca H and K break, rendering these stars particularly faint in
$B_T$.  

Therefore, we decided to investigate possible nonlinearities in the $B_T$ scale at
faint levels by plotting the ratios of $BVRI$ scale factors to those from the three space-based
bands against Hipparcos-Tycho magnitude, for all our Landolt cool giants and our 
Carter-Meadows A-dwarfs.  
Figures~\ref{bt},~\ref{vt}, and~\ref{hp} offer these plots for $B_T$, $V_T$, 
and $H_p$, respectively,
distinguishing between K-/M-giants and A-dwarfs.  On inspection, the A-stars never show any 
dependence of the ratio of scales on brightness, even in $B_T$, while the K-/M-stars present 
an obvious trend toward larger ratios at fainter magnitudes in $B_T$ but no similar
trends in the other two bands.

\begin{table}
\dummytable\label{htratio}
\end{table}

These trends are quantified in Table~\ref{htratio} where we give the inverse-variance weighted
mean ratios of ground-based to Hipparcos-Tycho scale factors for all the stars discussed in
Tables~\ref{stes} and ~\ref{stesavs}.  Table~\ref{htbias} presents the slopes of the regression
lines of scale-ratio against space-based magnitude.  The concurrence of these three ways of
examining the data on normalization ratios points to the conclusion that, for the faintest cool 
giants, one should not rely on the normalization factor derived directly from the $B_T$ band.  
If a faint cool giant has $BVRI$ and space-based data, then the influence of $B_T$ on the 
overall scale factor will be limited because the associated uncertainty in a faint 
$B_T$ magnitude will be large enough that the weight accorded to $B_T$'s contribution 
under inverse-variance weighting will be correspondingly small.  If a cool giant lacks any $UBV(RI)$
photometry, but Hipparcos-Tycho measurements exist, one could in principle ignore $B_T$'s 
scale factor.  However, this could reduce the number of optical normalizing wavelengths
to only a single, $V_T$, point or at best, two points, if $H_p$ is also available.
To handle such situations we suggest that it is important to use the regression line to
increase the $B_T$ scale factor, in accord with this line, so that it would
mimic the scale that would have been obtained using $BVRI$ observations.

The fact that A-dwarfs exhibit no such anomalies even in this bluest Hipparcos-Tycho
band argues for the effect to arise from the extremely different energy distributions
of the K-/M-giants, whose faintest optical region lies within $B_T$, and the A-stars,
which are brightest in the blue.  A simple test of this conjecture would be to seek a dependence 
of the $B_T$ scale factor on spectral type.  The set of spectral classes probed is K0, 1.5,
2, 3, 4, 5III, represented in our data set by 4, 1, 5, 1, 1, 1 stars, respectively.
A linear regression analysis using inverse-variance weighting for the scale factors
found at each spectral class indicates an offset of 1.12$\pm$0.02 (so that, even at type K0III,
the $B_T$ scale is not unity) and a significant slope of 0.04$\pm$0.01.  Figure~\ref{bt1kgt}
displays these data and the formal best fit regression line.

Since the time we began this program, the Tycho-2 catalog has appeared
(H{\o}g et al. 2000).  This is substantially larger than Tycho-1 (2.5 million stars
as compared with 1 million), and has higher quality photometry.  Therefore,
we decided to examine data from Tycho-2 in light of the trends found in Tycho-1
described above.  Figures~\ref{bt2} and~\ref{vt2} present the identical information 
to Figures~\ref{bt} and~\ref{vt} but 
based on Tycho-2 magnitudes.  This time the ratios of supertemplate scale factors
for the cool giants do not show any significant slope.  The formal slopes and offsets of
the regression lines against B$_T$ are 0.02$\pm$0.02 and 0.96$\pm$0.20
(for K-giants), and -0.03$\pm$0.05 and 1.27$\pm$0.44 (A-dwarfs).  Therefore,
we have replaced our Tycho-1 photometry for the stars in this paper by the
corresponding data from Tycho-2, whenever available. For 12 cool giants with both
Tycho-1 and Tycho-2 photometry we have investigated the effects of making this change
on the ratios of $BVRI$-determined scale factors to those from Tycho alone. In $B_T$, 
these ratios are worse (i.e. larger) with Tycho-2 than with Tycho-1 for 3 stars, unchanged
for 2 stars, but better for 7.  In $V_T$, where the effects are much smaller in magnitude,
the ratios are worse for 2 stars, unchanged for 5, and better for 5.  The regression
analysis for Tycho-2 B$_T$ magnitudes between ratios of scale factors and K spectral
class appears in Figure~\ref{bt2kgt} along with the formal best fit regression line.
For Tycho-2 there is still a significant offset (1.12$\pm$0.02) but the formal value
for the slope is insignificantly different from unity (0.01$\pm$0.01).  Therefore, we
consider it appropriate to rescale {\it all} K- and M-giant scale factors derived from 
Tycho-2 B$_T$ data by $\times$1.14$\pm$0.01 (corresponding to the inverse-variance
weighted average of the 12 K-giants' factors), thereby bringing them into alignment
with $BVRIH_pV_T$ normalization multipliers when we create absolute supertemplates. 
For those cool giants with Tycho-1 data but lacking Tycho-2 data (one star in our sample), 
we use the regression line in Figure~\ref{bt1kgt} to correct B$_T$ scale factors for Tycho-1 
photometry of K-/M-giants.

Advanced methods of data reduction, not available at the time Tycho-1 was processed, were 
applied to the Tycho-2 catalog (H{\o}g et al. 2000), for example, all data series for a given 
star throughout the entire mission were coadded with appropriate weighting prior to deriving 
photometry from the accumulated signal.  For low signal-to-noise signals in the Tycho 
star-mappers, the gain in Tycho-2 was substantial, compared with Tycho-1. Known inherent
biases in Tycho-1 are highlighted by H{\o}g et al. (2000).  Single transits of a star over
the star-mappers in which no obvious peak was identifiable were neglected in Tycho-1, leading to
preferential selection of brighter transits of the same star (\S4.1 of H{\o}g et al. 2000).
Second, in the presence of a changing background, the much longer series of sky samples
used in Tycho-1 can lead to systematic offsets in the assessed background level
(\S4.2 of H{\o}g et al. 2000). Both $V_T$
and $H_p$ sample these cool stellar energy distributions well past the 4000\AA\ break,
where their spectra rise rapidly into the red, so there are no corresponding problems
with cool stars in $V_T$ or $H_p$.
 
We suggest that the phenomenon we see is due to the already known 
biases in Tycho-1 but, at least partly, to Malmquist
bias whereby the detection of faint red stars in B$_T$ and, to a lesser extent in $V_T$,
is enhanced by upward noise fluctuations and 
this problem systematically worsens with later-type cool giants.
 Therefore, the
K-/M-giants appear brighter than reality, leading to smaller scale factors as
stars become fainter and cooler.  The improvements in processing of Tycho-2 alleviate
this problem, and avoid the known problem of the ``dramatic" ``faint end bias" intrinsic to 
Tycho-1 (\S7.1 of H{\o}g et al. 2000).  

\begin{table}
\dummytable\label{htbias}
\end{table}

\newpage

\begin{figure}
\caption{The IR template of a K4III derived by interpolation between the observed 
spectra of $\alpha$ Hya (K3III) and $\alpha$ Tau (K5III).  Solid lines represent
the spectra; short-dashed lines show the $\pm$1$\sigma$ bounds. }
\label{k4g}
\end{figure}

\begin{figure}
\caption{Comparison of the K4III template with KAO (filled squares: resolving
power $\sim$200) and LRS (open squares: resolving power 25--50) 
spectral observations of the K4 giant, $\beta$ UMi.  Solid line: K4III
template; dashed-dotted lines show the $\pm$2$\sigma$ bounds on the template.  The
observed spectral points carry $\pm$1$\sigma$ error bars.}
\label{kaocfn}
\end{figure}

\begin{figure}
\caption{The complete K5III supertemplate, combining Pickles' optical K5III spectrum 
with our observed infrared spectrum of $\alpha$ Tau.}
\label{ste}
\end{figure}

\begin{figure}
\caption{FAST spectra of giant stars with spectral types taken from
the catalog of Keenan \& McNeil (1989).  Although the entire 4000--7000\AA\ range 
was observed in a single exposure, the spectra are displayed in two sections: 
a) 4000--5500\AA\ and b) 5500--7000\AA.  These plots show a clear and pronounced
progression of numerous spectral features with spectral type.}
\label{mho1}
\end{figure}

\begin{figure}
\caption{FAST spectra of dwarf, giant and supergiant stars with spectral types ranging from G8 to M2.  
The luminosity classes and spectral types were taken from the catalog of Keenan \& McNeil (1989).
As in Figure~\ref{mho1}, the spectra are displayed in two plots spanning 4000--5500\AA ~(a) and 
5500--7000\AA ~(b). Both ranges show features with an obvious luminosity dependence;
in particular, the Ca~I line at 6122\AA\ and the Fe~II feature at 5169\AA.}
\label{mho2}
\end{figure}

\begin{figure}
\caption{The individual stars of our test sample with their MADs and 1$\sigma$
uncertainties.  Both A-dwarfs and cool giants are plotted together at each of the three
wavelengths: a) $J$; b) $H$; c) $K$.}
\label{nirmads}
\end{figure}

\begin{figure}
\caption{A montage of 8 supertemplates and models drawn from the 24 cool giants and 9
hot dwarfs discussed in this paper.  Mean curves (solid lines) are flanked by $\pm1\sigma$
uncertainty bounds (dashed lines).  The normalizing photometry is represented by filled circles with 
$\pm1\sigma$ error bars (sometimes within the circle symbols).}
\label{montage}
\end{figure}

\begin{figure}
\caption{The ratio of ground-based to space-based optical photometric normalization
factors for cool giants (filled squares) and A-dwarfs (open squares) against
Tycho-1's $B_T$ magnitude.}
\label{bt} 
\end{figure}

\begin{figure}
\caption{The ratio of ground-based to space-based optical photometric
normalization factors for cool giants (filled squares) and A-dwarfs (open
squares) against Tycho-1's $V_T$ magnitude.}
\label{vt}
\end{figure}

\begin{figure}
\caption{The ratio of ground-based to space-based optical photometric normalization
factors for cool giants (filled squares) and A-dwarfs (open squares) against $H_p$.}
\label{hp}
\end{figure}

\begin{figure}
\caption{The ratio of ground-based to Tycho-1 B$_T$ photometric normalization
factors for K-giants against K spectral class. The dashed line represents the formal
regression line, and the slope is statistically significant.}
\label{bt1kgt}
\end{figure}

\begin{figure}
\caption{The ratio of ground-based to space-based optical photometric normalization
factors for cool giants (filled squares) and A-dwarfs (open squares) against
Tycho-2's $B_T$ magnitude.}
\label{bt2} 
\end{figure}

\begin{figure}
\caption{The ratio of ground-based to space-based optical photometric
normalization factors for cool giants (filled squares) and A-dwarfs (open
squares) against Tycho-2's $V_T$ magnitude. The slope is statistically insignificant.}
\label{vt2}
\end{figure}

\begin{figure}
\caption{The ratio of ground-based to Tycho-2 B$_T$ photometric normalization
factors for K-giants against K spectral class. The slope is statistically insignificant.}
\label{bt2kgt}
\end{figure}  

\clearpage
\renewcommand{\arraystretch}{.6} 

\clearpage

\begin{deluxetable}{lllcrrl}
\centering
\tablewidth{0pt}
\tablenum{1}
\tablecolumns{7}
\tablecaption{ICRS 2000.0 coordinates for those Landolt SA stars and
found to be useful for our purposes. Source codes: ``S" - SIMBAD when queried 
by starname; ``G" - GSC1.3; ``M" - SIMBAD when queried by coordinates.
Spectral types are our own. \label{coords}} 
\tablehead{ Starname& RA& DEC& Source& $B$& $V$& Spectral type\\}
\startdata
SA92-336& 00 55 01.401& +00 47 22.39& S& 9.03& 8.05& K0III\\
SA94-251& 02 57 46.98&  +00 16 02.7& G& 12.45&11.22& K1III\\
SA103-526& 11 56 54.19&  --00 30 13.8& G& 11.99&10.90& K0III\\
SA105-205& 13 35 52.598& --00 57 53.56& S& 10.16& 8.80& K3III\\
SA105-405& 13 35 59.535& --00 34 39.62& S& 9.83& 8.31& K5III\\
SA107-35&  15 37 28.856& --00 53 05.44& S& 9.05& 7.78& K2III\\
SA107-347& 15 38 35.773& --00 35 57.74& S& 10.74& 9.44& K1.5III\\
SA107-484& 15 40 16.80&  --00 21 14.5& G& 12.55& 11.31& K3III\\
SA108-475& 16 37 00.09&  --00 34 40.0& S& 12.69& 11.31& K3III\\
SA108-827& 16 37 21.171& --00 24 48.59& S& 9.27& 7.96& K2III\\
SA108-1918& 16 37 50.12& --00 00 36.1& G& 12.82& 11.38& K2/3III\\
SA109-231& 17 45 19.965& --00 25 51.60& S& 10.79& 9.33& K2III\\
SA110-471& 18 41 26.680& +00 33 51.88& S& 8.92& 7.47& K2/3III\\
SA112-275& 20 42 35.426& +00 07 20.23& S& 11.12& 9.91& K0III\\
SA112-595& 20 41 18.47&  +00 16 28.3& G& 12.95& 11.35& M0III\\
SA113-259& 21 41 44.83&  +00 17 40.2& G& 12.94& 11.74& K2III\\
SA113-269& 21 42 01.322& +00 17 45.22& S& 10.59& 9.48& K0III\\
SA114-176& 22 43 10.183& +00 21 15.53& S& 10.72& 9.24& K4/5III\\
SA114-548& 22 41 36.85&  +00 59 06.1& G& 12.96& 11.60& K2.5III\\
SA114-656& 22 41 35.2&   +01 11 10&S& 13.61& 12.64& K1III\\
SA114-670& 22 42 09.310& +01 10 16.78& M& 12.31& 11.10& K1/2III\\
SA115-427& 23 43 14.430& +01 06 47.01 & S& 10.03& 8.86& K2III\\
SA115-516& 23 44 15.374& +01 14 12.65& S& 11.46& 10.43& K1.5III\\
HD197806& 20 47 30.773& --43 10 33.63& S& 10.84& 9.66& K0III\\
\enddata
\end{deluxetable}

\clearpage

\begin{deluxetable}{lllllll}
\rotate 
\centering
\tablewidth{0pt}
\tablenum{2}
\tablecolumns{7}
\tablecaption{Comparison of Landolt and Carter-Meadows $UBVRI$ data 
for two stars in common \label{cfns} }
\tablehead{ Star& System& $U\pm$unc.& $B\pm$unc.& $V\pm$unc.& $R\pm$unc.& $I\pm$unc.\\}
\startdata
SA94-251& Landolt&        13.704$\pm$0.0036& 12.423$\pm$0.0015& 11.204$\pm$0.0010& 10.545$\pm$0.0013&  9.957$\pm$0.0015\\
SA94-251& Carter-Meadows& 13.750$\pm$0.054&  12.454$\pm$0.016&  11.224$\pm$0.006&  10.572$\pm$0.009&   9.986$\pm$0.009\\
SA108-475& Landolt& 14.151$\pm$0.005& 12.689$\pm$0.002& 11.309$\pm$0.0014& 10.565$\pm$0.001& 9.900$\pm$0.002\\
SA108-475& Carter-Meadows& 14.165$\pm$0.020& 12.703$\pm$0.020& 11.314$\pm$0.013&  10.580$\pm$0.018& 9.916$\pm$0.016\\
\enddata
\end{deluxetable}

\clearpage
\begin{deluxetable}{cccccccc}  
\rotate
\centering
\tablewidth{0pt}
\tablenum{3}
\tablecolumns{8}
\tablecaption{Absolute calibration of the ground-based optical photometry systems supporting this work.
Quantities tabulated correspond to the definition of zero magnitude in each filter \label{gndcal} }
\tablehead{System/Band& In-band Flux& In-band Unc.& Bandwidth& F$_\lambda$(iso)& $\lambda$(iso)& F$_{\nu}$& AB$_{\nu}$ \\  
 & W cm$^{-2}$& W cm$^{-2}$&   $\mu $m&   W cm$^{-2}$ $\mu $m$^{-1}$& $\mu$m& Jy& m\\}
\startdata
Landolt-$U$& 2.823E-13& 4.353E-15& 0.0711& 3.971E-12& 0.3745& 1649& +0.857\\
Landolt-$B$& 5.371E-13& 8.242E-15& 0.0819& 6.562E-12& 0.4481& 4060& --0.121\\
Landolt-$V$& 3.327E-13& 5.086E-15& 0.0878& 3.789E-12& 0.5423& 3723& --0.027\\
Landolt-$R$& 3.860E-13& 5.772E-15& 0.1697& 2.274E-12& 0.6441& 3168& +0.148\\
Landolt-$I$& 1.438E-13& 2.159E-15& 0.1274& 1.129E-12& 0.8071& 2459& +0.423\\
\enddata  
\end{deluxetable}

\clearpage

\begin{deluxetable}{llccccccc} 
\centering
\tablewidth{0pt}
\tablenum{4}
\tablecolumns{9}
\tablecaption{New TCS $JHK$ photometry, and uncertainties, for Landolt stars \label{tcs} }
\tablehead{ Star& Date& $J$& $H$& $K$& $\epsilon$$J$& $\epsilon$$H$& $\epsilon$$K$& \# chop \\
& & m& m& m& m& m& m& cycles \\}
\startdata
SA92-336& 1999Jul12-14&    6.325& 5.828& 5.741& 0.014& 0.016& 0.016& 4\\
SA103-526&  1998May23-24& 8.993&     8.467&     8.372& 0.015&     0.008&     0.011& 7\\
SA105-205& 1998May23-24&  6.320&     5.628&     5.491& 0.013&   0.008&    0.006& 6\\
SA105-405& 1998May23-24&  5.532&    4.793&    4.626& 0.017&     0.011&  0.011& 6\\
SA107-35& 1998May23-24&   5.527&     4.946&     4.812& 0.007&     0.002&     0.002&   6\\
SA107-35& 1999Jul12-14 &   5.515& 4.932& 4.804& 0.003& 0.009& 0.006& 4 \\
SA107-35& Combined &   5.517& 4.945& 4.811& 0.003& 0.002& 0.002& 10 \\
SA107-347& 1998May23-24&  7.004&  6.322& 6.198& 0.003&     0.004&     0.006& 4\\
SA107-347& 1999Jul12-14& 7.025& 6.338& 6.207& 0.016& 0.008& 0.008&   4  \\
SA107-347& Combined & 7.005&   6.325&   6.201&  0.003&   0.004&   0.005& 8\\
SA107-484&  1998May23-24& 9.097&     8.516&     8.393& 0.023&     0.010&     0.010& 6\\
SA107-484&  1999Jul12-14&   9.133& 8.509& 8.392& 0.047& 0.012& 0.024& 5\\
SA107-484&  Combined&   9.104& 8.513& 8.393& 0.021& 0.008& 0.009& 11\\
SA108-475& 1998May23-24&   8.789& 8.105& 7.965& 0.024&     0.010&     0.009& 4\\
SA108-475& 1999Jul12-14 &   8.761& 8.108& 7.942& 0.023& 0.027& 0.020& 5\\
SA108-475& Combined&   8.774& 8.105& 7.961& 0.017& 0.009& 0.008& 9\\
SA108-827&  1998May23-24&  5.685&     5.104&     4.970&  0.016&     0.003&     0.003& 4\\
SA108-827& 1999Jul12-14 &   5.679& 5.092& 4.958& 0.005& 0.009& 0.006& 4\\
SA108-827& Combined&   5.680& 5.103& 4.968& 0.005& 0.003& 0.003& 8\\
SA108-1918& 1998May23-24&  8.816&     8.147&     8.000& 0.022&     0.007&    0.006& 4\\
SA108-1918& 1999Jul12-14 &  8.814& 8.152& 8.003& 0.049& 0.015& 0.015 & 6\\
SA108-1918& Combined&  8.816& 8.148& 8.000& 0.020& 0.006& 0.006 & 10\\
SA109-231&   1998May23-24&  6.659&     6.031&     5.866& 0.020&     0.012&     0.011& 4\\
SA109-231& 1999Jul12-14 &   6.669& 6.036& 5.868& 0.005& 0.006& 0.003& 6    \\
SA109-231& Combined&   6.668& 6.035& 5.868& 0.005& 0.005& 0.003& 10    \\
SA110-471&  1998May23-24&   4.918&     4.236&     4.07& 0.006&     0.011&     0.005& 4\\
SA110-471& 1999Jul12-14 &   4.931& 4.244& 4.088& 0.002& 0.002& 0.002 & 4   \\
SA110-471& Combined&   4.930& 4.244& 4.086& 0.002& 0.002& 0.002 & 8   \\
SA112-275&   1998May23-24&  7.751&     7.164&     7.047& 0.005&     0.003&     0.005& 2\\
SA112-275& 1999Jul12-14 &   7.765& 7.176& 7.063& 0.022& 0.007& 0.006& 4\\
SA112-275& Combined&   7.752& 7.166& 7.054& 0.005& 0.003& 0.004& 6\\
SA112-595&   1998May23-24& 8.287&     7.470&     7.280& 0.006&     0.008&     0.007& 4\\
SA112-595& 1999Jul12-14 &   8.312& 7.480& 7.304& 0.030& 0.012& 0.017& 5   \\
SA112-595& Combined&   8.288& 7.473& 7.283& 0.006& 0.007& 0.006& 9   \\
SA113-259& 1999Jul12-14 &   9.702& 9.120& 8.986& 0.111& 0.033& 0.057 & 9   \\
SA113-269& 1999Jul12-14 &   7.564& 7.022& 6.918& 0.012& 0.007& 0.007& 6    \\
SA114-176& 1999Jul12-14&   6.616& 5.914& 5.752& 0.011& 0.005& 0.009& 4  \\
SA114-548& 1999Jul12-14&   9.176& 8.500& 8.333& 0.034& 0.024& 0.030& 4  \\
SA114-656& 1999Jul12-14&  10.814&    10.211&    10.209& 0.105& 0.097& 0.090& 6\\ 
SA114-670& 1999Jul12-14 &   8.964& 8.374& 8.264& 0.036& 0.011& 0.039& 7  \\
SA115-427& 1999Jul12-14 &   6.794& 6.212& 6.107& 0.013& 0.008& 0.010 &6   \\
SA115-516& 1999Jul12-14 &   8.488& 7.926& 7.836& 0.020& 0.007& 0.015& 4    \\
\enddata
\end{deluxetable}

\clearpage
\begin{deluxetable}{llll}
\tablewidth{0pt}
\tablenum{5}
\tablecolumns{4}
\tablecaption{The new technique of optical normalization of the supertemplate
   for a Landolt K-giant, SA112-275, K0III, A$_V$=0.620, using just Landolt's
$UBVRI$ photometry \label{112275opt} }
\tablehead{ Filter& Scale& Frac.Unc.& $\lambda_{iso}$($\mu$m)}
\startdata
  LU& 3.995E-04& 5.360E-03&   0.3597\\
  LB& 6.029E-04& 5.190E-03&   0.4409\\
  LV& 5.988E-04& 4.853E-03&   0.5440\\
  LR& 6.158E-04& 4.660E-03&   0.6427\\
  LI& 5.333E-04& 4.827E-03&   0.8048\\
Mean scale& 5.554E-04& 2.218E-03& \nodata \\
\enddata
\end{deluxetable}

\clearpage
\begin{deluxetable}{cccccc} 
\centering
\tablewidth{0pt}
\tablenum{6}
\tablecolumns{6}
\tablecaption{Predicting Tenerife $JHK$ from the final, optically normalized, reddened 
supertemplate of SA112-275 \label{112275tcs} }
\tablehead{ &  \multicolumn{2}{c}{Predicted}& \multicolumn{2}{c}{Observed} \\
Filter&  Magnitude& Uncertainty& Magnitude& Uncertainty& Difference\\}
\startdata
TCSJ&   7.706&  0.016& 7.752& 0.005&  +0.046\\
TCSH&   7.191&  0.011& 7.166& 0.003& --0.025\\
TCSK&   7.083&  0.010& 7.054& 0.004& --0.029\\
MAD&  \nodata&  \nodata&  \nodata&  \nodata& --0.003\\
\enddata
\end{deluxetable}

\clearpage
\begin{deluxetable}{llllrrl}
\centering
\tablewidth{0pt}
\tablenum{7}
\tablecolumns{7}
\tablecaption{Cool giant supertemplates: spectral types, A$_V$s, and 
final scale factors and biases, derived by normalizing using all available optical, NIR,
and MIR photometry; near-infrared ($JHK$) MADs derived from from the optical 
photometry listed in the final column \label{stes} }
\tablehead{ Star&  Template& A$_V$& Scale&   Bias(\%)& MAD& Photometry sources\\}
\startdata 
SA92-336&   K0III&   0.000& 1.775E-03& 0.30& +0.006& Landolt/TCS\\
SA94-251&   K1III&   0.493& 2.969E-05& 0.33& --0.014& Carter-Meadows/SAAO\\
SA103-526&  K0III&   0.245& 1.585E-04& 0.32& --0.021& Landolt/TCS\\
SA105-205&  K3III&   0.353& 2.063E-03& 0.32& --0.028& Landolt/TCS\\
SA105-405&  K5III&   0.031& 9.528E-04& 0.33& -0.044& Landolt/TCS\\
SA107-35&   K2III&   0.263& 3.943E-03& 0.29& --0.029& Landolt/TCS\\
SA107-347&  K1.5III&  0.608& 2.011E-04& 0.31& --0.062& Landolt/TCS\\
SA107-484&  K3III&   0.000& 1.445E-04&  0.38& +0.018& Landolt/TCS\\
SA108-475&  K3III&   0.406& 2.110E-04& 0.39& --0.042& Landolt/TCS\\
SA108-475&  K3III&   0.434& 2.156E-04& 0.92& +0.008& Carter-Meadows/SAAO\\
SA108-475&  K3III&   0.422& 2.143E-04& 0.33& \nodata& All optical+NIR\\
SA108-827&  K2III&   0.350& 3.558E-03& 0.32& +0.008& Landolt/TCS\\
SA108-1918& K3III&   0.567& 2.237E-04& 0.39& +0.039& Landolt/TCS\\
SA109-231&  K2III&   0.843& 1.611E-03& 0.32& --0.011& Landolt/TCS\\
SA110-471&  K2III&   0.803& 8.465E-03& 0.31& +0.036& Landolt/TCS\\
SA112-275&  K0III&   0.620& 5.628E-04& 0.30& --0.003& Landolt/TCS\\
SA112-595&  M0III&   0.174& 2.098E-04& 0.35& --0.030& Landolt/TCS\\
SA113-259&  K2III&   0.012& 8.157E-05& 0.31& --0.008& Landolt/TCS\\
SA113-269&  K0III&   0.316& 6.245E-04& 0.32& +0.015& Landolt/TCS\\
SA114-176&  K4III&   0.046& 3.388E-04& 0.29& --0.014& Landolt/TCS\\
SA114-548&  K3III&   0.350& 1.534E-04& 0.35& +0.012& Landolt/TCS\\ 
SA114-656&  K1III&   0.000& 5.062E-06& 0.43& --0.016& Landolt/TCS\\ 
SA114-670&  K1.5III&   0.329& 3.267E-04& 0.34& +0.045& Landolt/TCS\\
SA115-427&  K2III&   0.000& 1.159E-03& 0.30& --0.025& Landolt/TCS\\
SA115-516&  K1.5III&   0.000& 4.508E-05& 0.30& +0.005& Landolt/TCS\\
HD197806&   K0III&   0.530& 6.526E-04& 0.53& --0.013& Carter-Meadows/SAAO\\
\enddata
\end{deluxetable}

\clearpage
\begin{deluxetable}{llllrrl}
\centering
\tablewidth{0pt}
\tablenum{8}
\tablecolumns{7}
\tablecaption{Carter-Meadows A-dwarfs: spectral types, A$_V$s, and 
final scale factors and biases, derived by normalizing using all available optical, NIR,
and MIR photometry; near-infrared ($JHK$) MADs derived from from the optical 
photometry listed in the final column  \label{stesavs} }
\tablehead{ Star&  Template& A$_V$ & Scale&   Bias(\%)& MAD& Photometry sources\\}
\startdata
HD15911&  A0V&  0.000& 1.272E+02& 1.14& +0.005& Carter-Meadows/SAAO\\
HD29250&  A4V&  0.295& 1.699E+02& 1.10& +0.023& Carter-Meadows/SAAO\\ 
HD62388&  A0V&  0.232& 2.637E+02& 1.12& -0.028& Carter-Meadows/SAAO\\
HD71264&  A0V&  0.406& 3.152E+02& 1.16& +0.029& Carter-Meadows/SAAO\\
HD84090&  A3V&  0.149& 3.532E+02& 1.13& +0.009& Carter-Meadows/SAAO\\
HD105116& A2V&  0.372& 5.138E+02& 1.12& -0.053& Carter-Meadows/SAAO\\
HD106807& A1V&  0.276& 2.850E+02& 1.09& -0.035& Carter-Meadows/SAAO\\
HD136879& A0V&  0.521& 3.124E+02& 1.13& -0.006& Carter-Meadows/SAAO\\
HD216009& A0V&  0.155& 5.171E+02& 1.08& -0.008& Carter-Meadows/SAAO\\
\enddata
\end{deluxetable}

\clearpage

\begin{deluxetable}{lllcccccccc}
\rotate 
\centering
\tabletypesize
\small
\tablewidth{0pt}
\tablenum{9}
\tablecolumns{11}
\tablecaption{Predicted and observed IRAS flux densities for the reddened
supertemplates scaled exactly as for $JHK$ predictions \label{iras} }
\tablehead{& & & \multicolumn{4}{c}{PREDICTED}& \multicolumn{4}{c}{OBSERVED}\\
Star& PSC& FSC/R& PSC& FSC/R& PSC& FSC/R& PSC& FSC/R& PSC& FSC/R\\
& & & (12 $\mu$m)& (12 $\mu$m)& (25 $\mu$m)& (25 $\mu$m)& (12 $\mu$m)& (12 $\mu$m)& (25 $\mu$m)& (25 $\mu$m)\\
& & & [mJy]& [mJy]& [mJy]& [mJy]& [mJy]& [mJy]& [mJy]& [mJy]\\}
\startdata
SA92-336& \nodata& F00525+0031& 223$\pm$4&  222$\pm$4&  52$\pm$1& 51$\pm$1& \nodata& 214$\pm$32& \nodata& $<$241\\
SA105-405& 13334--0019& F13333--0019&  619$\pm$9&  615$\pm$9&   152$\pm$1&  151$\pm$1& 650$\pm$59& 671$\pm$60& $<$290&  $<$279\\
SA107-35& 15349--0043& F15349--0043&  528$\pm$11&  525$\pm$10&   129$\pm$2&  128$\pm$2& 470$\pm$33& 507$\pm$30& $<$300&  $<$228\\
SA107-347& \nodata& Z15360--0026& 143$\pm$2&  142$\pm$2&    36$\pm$1&   35$\pm$1& \nodata& 143$\pm$29& \nodata& $<$101 \\
SA108-827& 16347--0018& F16347--0018&  477$\pm$10&  474$\pm$9&   117$\pm$2&  116$\pm$2&  470$\pm$56& 465$\pm$47& $<$280& $<$185 \\
SA109-231& \nodata& F17247--0024&  210$\pm$4&  209$\pm$4&    53$\pm$2&   52$\pm$2& \nodata& 173$\pm$29& \nodata& $<$91\\
SA110-471& \nodata& Z18389+0030& 1114$\pm$23& 1108$\pm$22&   279$\pm$5&  276$\pm$5&  \nodata&  1030$\pm$134& \nodata& $<$750\\
SA113-269& \nodata& Z21394+0003& 77$\pm$1&   77$\pm$1&    18$\pm$1&   18$\pm$1&  \nodata& 92$\pm$25& \nodata& $<$151\\
SA115-427& \nodata& Z23406+0050& 158$\pm$3&  157$\pm$3&    38$\pm$1&   38$\pm$1&  \nodata& 153$\pm$41& \nodata& $<$314\\
\enddata
\end{deluxetable}

\clearpage

\begin{deluxetable}{ccccc} 
\centering
\tablewidth{0pt}
\tablenum{10}
\tablecolumns{5}
\tablecaption{Predicted and observed MSX 8.28-$\mu$m attributes (with
uncertainties) for the reddened supertemplate of SA114-176 scaled exactly as 
for $JHK$ predictions [\#1] and again using all of $BVRI$$B_T$$V_T$$JHK$ [\#2] \label{msx} }
\tablehead{Photometry& Irradiance& F$_{\lambda}$(iso)& [8.28]& F$_{\nu}$\\
\nodata& 10$^{-18}$ W cm$^{-2}$~$\mu$m$^{-1}$& 10$^{-18}$ W cm$^{-2}$~$\mu$m$^{-1}$& m& mJy\\}
\startdata 
Observed&       3.81$\pm$0.24& 1.13$\pm$0.07& 5.83$\pm$0.07& 272$\pm$17\\
Predicted[\#1]& 4.43$\pm$0.22& 1.32$\pm$0.09& 5.67$\pm$0.06& 316$\pm$16\\
Predicted[\#2]& 4.43$\pm$0.22& 1.32$\pm$0.09& 5.67$\pm$0.06& 316$\pm$16\\
\enddata
\end{deluxetable}

\clearpage
\begin{deluxetable}{cccrcccccc} 
\centering
\rotate 
\tablewidth{0pt}
\tablenum{11}
\tablecolumns{10}
\tablecaption{Expected IRAC attributes and uncertainties for the two faintest
stars in our sample of Landolt/Carter-Meadows objects \label{goals}}
\tablehead{Star&     Type&  Band&   Mag.&    Mag.unc& F$_{\lambda}$& unc.F$_{\lambda}$&
F$_{\nu}$(iso)& unc.F$_{\nu}$(iso)& Frac.unc.\\
& & & m& m& W cm$^{-2}$~$\mu$m$^{-1}$& W cm$^{-2}$~$\mu$m$^{-1}$& mJy& mJy&  \% \\}
\startdata
     HD15911& A0V& IRAC1&  9.470&  0.023& 1.07E-18& 2.08E-20& 45.1& 0.88& 1.94\\
     \nodata& \nodata& IRAC2&  9.470&  0.023& 4.32E-19& 8.38E-21& 29.2& 0.57& 1.94\\
     \nodata& \nodata& IRAC3& 9.469&  0.023& 1.77E-19& 3.42E-21& 19.0& 0.37& 1.94\\
     \nodata& \nodata&  IRAC4& 9.472& 0.023& 4.90E-20& 9.46E-22& 10.2& 0.20& 1.93\\
   SA114-656& K1III& IRAC1& 10.045&  0.023& 6.10E-19& 1.28E-20& 25.7& 0.54& 2.10\\
   \nodata& \nodata& IRAC2& 10.155&  0.030& 2.22E-19& 6.19E-21& 15.0& 0.42& 2.78\\
   \nodata& \nodata& IRAC3& 10.117&  0.018& 9.41E-20&1.56E-21& 10.1& 0.17& 1.66\\
   \nodata& \nodata& IRAC4& 10.063&  0.019& 2.75E-20& 4.96E-22& 5.75E+00& 0.10& 1.80\\
\enddata
\end{deluxetable}

\clearpage
\begin{deluxetable}{cccccccc} 
\rotate
\centering
\tablewidth{0pt}
\tablenum{12}
\tablecolumns{8}
\tablecaption{Absolute calibration of the Hipparcos-Tycho optical photometry systems 
supporting this work. Quantities tabulated correspond to the definition of zero magnitude 
in each filter  \label{htcal} }
\tablehead{ System/Filter& In-band Flux& In-band Unc.& Bandwidth& F$_{\lambda}$(iso)& $\lambda$(iso)& 
F$_{\nu}$(iso)& AB$_{\nu}$ \\
\nodata& W cm$^{-2}$& W cm$^{-2}$&   $\mu$m&   W cm$^{-2}$$\mu$m$^{-1}$& $\mu$m& Jy \\}
\startdata
Tycho-$B_T$& 4.595E-13& 8.656E-15& 0.0685& 6.714E-12& 0.4394& 3943& --0.090\\
Tycho-$V_T$& 4.156E-13& 8.669E-15& 0.1033& 4.025E-12& 0.5323& 3761& --0.038\\
Hipparcos-$H_p$& 9.412E-13& 1.941E-14& 0.2383& 3.950E-12& 0.5355& 3748& --0.034\\
\enddata
\end{deluxetable}

\clearpage
\begin{deluxetable}{cccc}
\centering
\tablewidth{0pt}
\tablenum{13}
\tablecolumns{4}
\tablecaption{Ground-based supertemplate normalization compared with Hipparcos/Tycho-1 scale
factors, as a function of space-based brightness: inverse-variance weighted
mean ratios and $\sigma$s of scale factors \label{htratio} }
\tablehead{Filter&  K-/M-giants&  A-dwarfs& All stars\\}
\startdata
$B_T$&   1.180$\pm$0.009& 0.994$\pm$0.024& 1.155$\pm$0.009\\
$V_T$&   1.012$\pm$0.004& 0.997$\pm$0.014& 1.011$\pm$0.004\\
$H_p$&   1.032$\pm$0.011& 0.998$\pm$0.029& 1.030$\pm$0.011\\
\enddata
\end{deluxetable}

\clearpage
\begin{deluxetable}{cccc}
\centering
\tablewidth{0pt}
\tablenum{14}
\tablecolumns{4}
\tablecaption{Slopes of the regression lines of (ground-based/space-based) scale factors
against space-based magnitudes, in the three Hipparcos-Tycho bands, based on
Tycho-1 and Tycho-2 photometry  \label{htbias} }
\tablehead{Filter&  K-/M-giants&  A-dwarfs& All stars\\}
\startdata
Tycho-1& & & \\
$B_T$&   +0.033$\pm$0.018& --0.036$\pm$0.052& +0.062$\pm$0.016\\
$V_T$&   +0.020$\pm$0.006& --0.007$\pm$0.025& +0.015$\pm$0.006\\
$H_p$&   +0.061$\pm$0.026&  +0.001$\pm$0.031& +0.037$\pm$0.026\\
Tycho-2& & & \\
$B_T$&   +0.018$\pm$0.019& --0.028$\pm$0.048& +0.034$\pm$0.015\\
$V_T$&   +0.015$\pm$0.010& --0.008$\pm$0.026& +0.011$\pm$0.010\\
$H_p$&   +0.015$\pm$0.019& --0.003$\pm$0.029& +0.015$\pm$0.024\\
\enddata
\end{deluxetable}


\newpage

\begin{thebibliography}{48}
\bibitem{} Bessell, M.S. 1979, PASP, 91, 589
\bibitem{} Bessell, M.S. 1983, PASP, 95, 480
\bibitem{} Bessell, M.S. 1990, PASP, 102, 1181
\bibitem{} Bessell, M.S., Castelli, F., \& Plez, B. 1998, A\&A, 333, 231
\bibitem{} Bessell, M.S. 2000, PASP, 112, 961
\bibitem{} Cardelli, J.A., Clayton, G.C., \& Mathis, J.S. 1989, \apj, 345, 245
\bibitem{} Carter, B.S. \& Meadows, V.S. 1995, MNRAS, 276, 734
\bibitem{} Cohen, M. 1993, \aj, 105, 1860
\bibitem{} Cohen, M. 1998, \aj, 115, 2092
\bibitem{} Cohen, M., Walker, R.G., Jayaraman, S., Barker, E., \& Price, S.D. 2001, AJ, 121, 1180
\bibitem{} Cohen, M., Hammersley, P.L., \& Egan, M.P. 2000, AJ, 120, 3362
\bibitem{} Cohen, M., Walker, R. G., Barlow, M. J. \& Deacon, J. R. 1992b, \aj, 104, 1650
\bibitem{} Cohen, M., Walker, R. G., Carter, B., Hammersley, P. L., Kidger, M. R., \& Noguchi, K. 1999, \aj, 117, 1864 (Paper X)
\bibitem{} Cohen, M., Walker, R. G., \& Witteborn, F. C. 1992a, \aj, 104, 2030
\bibitem{} Cohen, M., Wheaton, W., Skrutskie, M.F., Megeath, S.T., \& Cutri, R. 2003, in preparation
\bibitem{} Cohen, M., Witteborn, F. C., Walker, R. G., Bregman, J. D., \& Wooden, D. H. 1995, AJ, 110, 275 
\bibitem{} Cohen, M., Witteborn, F. C., Bregman, J. D., Wooden, D. H., Salama, A., \& Metcalfe, L. 1996a, AJ, 112, 241
\bibitem{} Cohen, M., Witteborn, F. C., Carbon, D. F., Davies, J. K., Wooden, D. H. \& Bregman, J. D. 1996b, AJ, 112, 2274
\bibitem{} Cousins, A.W.J. 1984, S. Afr. astr. Obs. Circ., 1, no. 8, 69
\bibitem{} Drilling, J.S. \& Landolt, A.U. 1971, \aj, 84, 783
\bibitem[]{} Egan, M.P. et al. 1999, ``The Midcourse Space Experiment Point Source Catalog Version 1.2 Explanatory Guide'', Air Force Research Laboratory Technical Report, AFRL-VS-TR 1999-1522
\bibitem{} ESA, 1997, The Hipparcos and Tycho Catalogues, SP-1200 
\bibitem{} Fabricant, D., Cheimets, P., Caldwell, N., \& Geary, J. 1998, PASP, 110, 79
\bibitem{} Fitzgerald, M. P. 1970, A\&A, 4, 234
\bibitem{} Fitzpatrick, E.L. 1999, PASP, 111, 63
\bibitem{} Fluks, M. A. et al. 1994, A\&AS, 105, 311
\bibitem{} Hauser, M.G. et al. 1998, ApJ, 508, 25
\bibitem{} Hayes, D.S. \& Latham, D.W. 1975, ApJ, 197, 593 
\bibitem{} H{\o}g, E. et al. 1997, A\&A, 323, L57
\bibitem{} H{\o}g, E. et al. 2000, A\&A, 335, 367
\bibitem{} Houk, N. \& Swift, C. 1999, ``Michigan Catalogue of two-dimensional 
spectral types for the HD Stars, vol. 5" (Ann Arbor, Michigan : Department
of Astronomy, University of Michigan) 1999
\bibitem{} de Jager, C. \& Nieuwenhuijzen, H. 1987, A\&A, 177, 217
\bibitem{} Keenan, P.C. \& McNeil, R.C. 1989, ApJS, 71, 245
\bibitem{} Kessler, M. F. et al. 1996, A\&A,, 315, L27
\bibitem{} Kurucz, R.L. 1993, CD-ROM series; ``ATLAS9 Stellar Atmosphere Programs and 2 km/s grid",
CD-ROM No.13); ``SYNTHE Spectrum Synthesis Programs and Line Data" (Kurucz CD-ROM No.118)
\bibitem{} Landolt, A.U. 1973, \aj, 78, 959
\bibitem{} Landolt, A.U. 1983, \aj, 88, 439
\bibitem{} Landolt, A.U. 1992, \aj,104, 340
\bibitem{} van Leeuwen, F. 1997, Sp. Sci. Rev., 81, 201
\bibitem{} Menzies, J.W. 1989, MNRAS, 237, 21P
\bibitem{} Menzies, J.W., Marang, F., Laing, J.D., Coulson, I.M., \& Engelbrecht, C.A. 1991, MNRAS, 248, 642
\bibitem{} Mermilliod, J.-C. 1993, priv.comm.
\bibitem{} Mill, J., O'Neil, R. R., Price, S. D., Romick, G. J., Uy, O. M., \& Gaposchkin, E. M. 1994, J.Spacecr. Rockets, 31, 900
\bibitem{} Murakami, H. et al. 1996, PASJ, 48, L41
\bibitem{} O'Donnell, J.E. 1994, \apj, 422, 158
\bibitem{} Oke, J.B. \& Gunn, J.E. 1983, ApJ, 266, 713
\bibitem{} Perry, C.L. \& Johnston, L. 1982, A\&AS 50, 451
\bibitem{} Perry, C.L., Johnston, L. \& Crawford, D.L.1982, AJ 87, 1751
\bibitem{} Pickles, A.J. 1998, PASP, 110, 863
\bibitem[]{} Price, S.D., Egan, M.P., Carey, S.J., Mizuno, D. \& Kuchar, T. 2001, AJ, 121, 2819
\bibitem{} Rieke, G.H. \& Lebofsky, M.J. 1985, \apj, 288, 618
\bibitem{} Straizys, V. 1992, ``Multicolor stellar photometry" (Pachart Publishing House: 
Tucson, AZ), Astronomy and astrophysics series, v. 15.
\bibitem{} Witteborn, F.C., Cohen, M., Bregman, J.D., Heere, K.R., Greene, T.P.,
Wooden, D.H. 1995, ASP Conf. Ser., 73, ``Airborne Astronomy Symposium on the 
Galactic Ecosystem: From Gas to Stars to Dust", eds. Michael R. Haas, Jacqueline 
A. Davidson, Edwin F. Erickson (ASP: San Francisco, CA), p. 573
\bibitem{} Walker, R.G. \& Cohen, M. 2002, ``Walker-Cohen Atlas of Calibrated 
Spectra, Explanatory Supplement to Release 4.0", Contractor's Report to 
USAF Phillips Laboratory, Contract F19628-98-C-0047
\end{thebibliography}
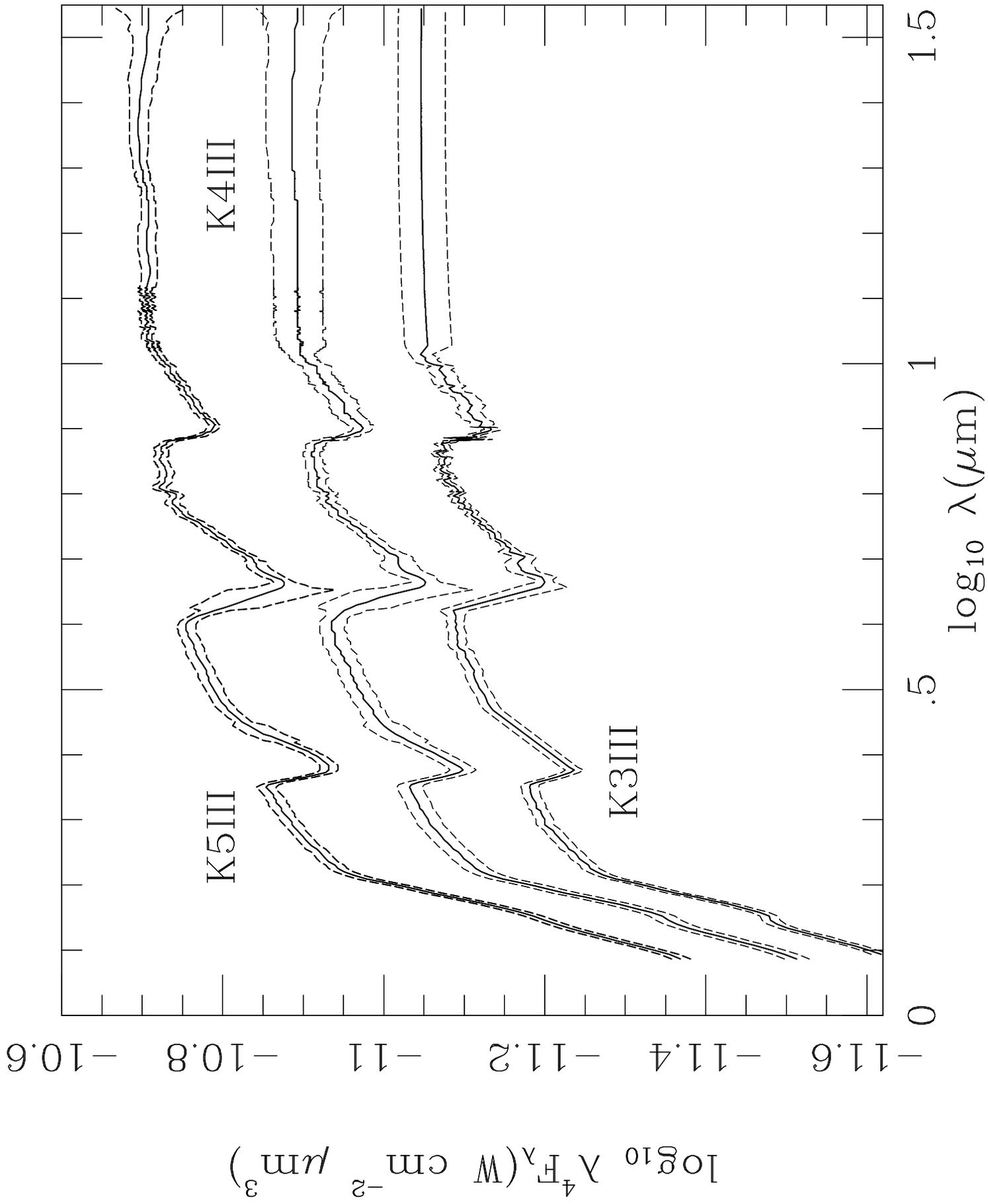
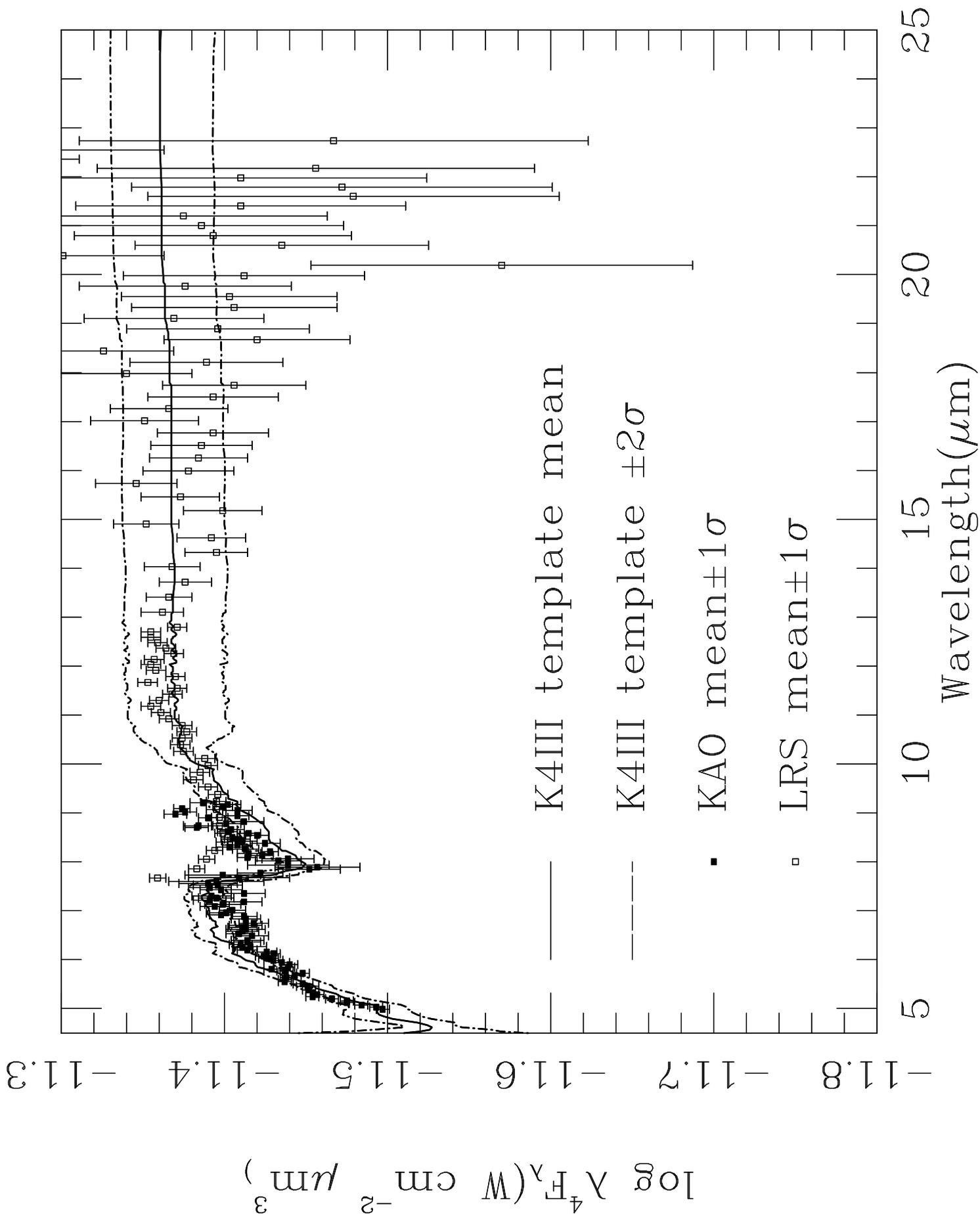
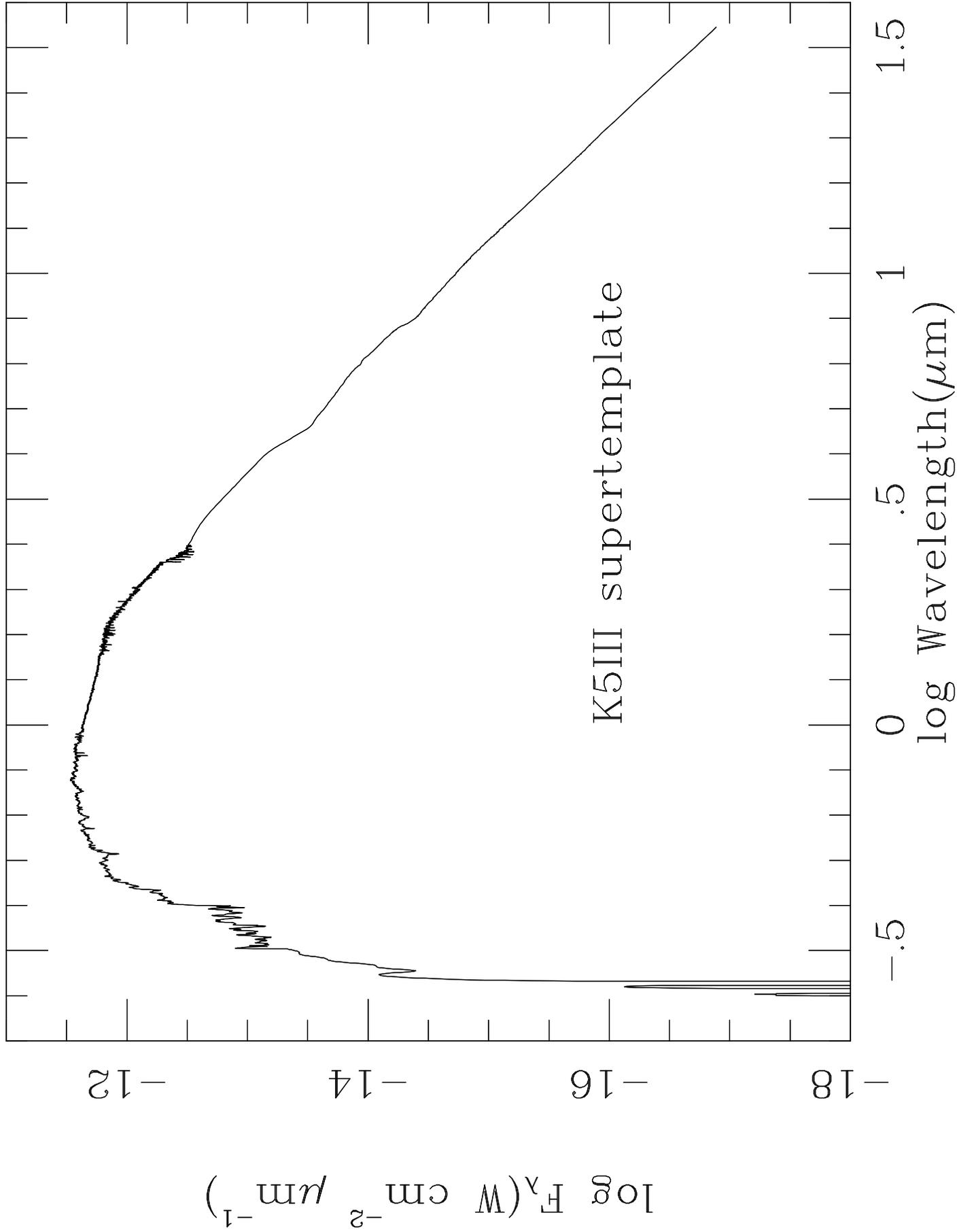
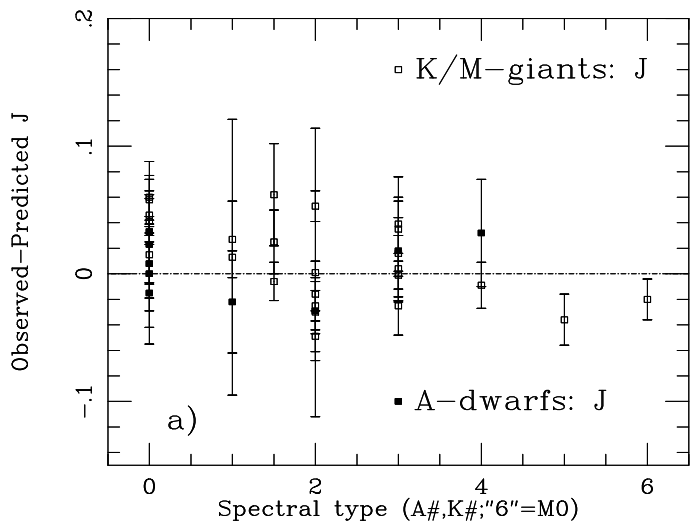
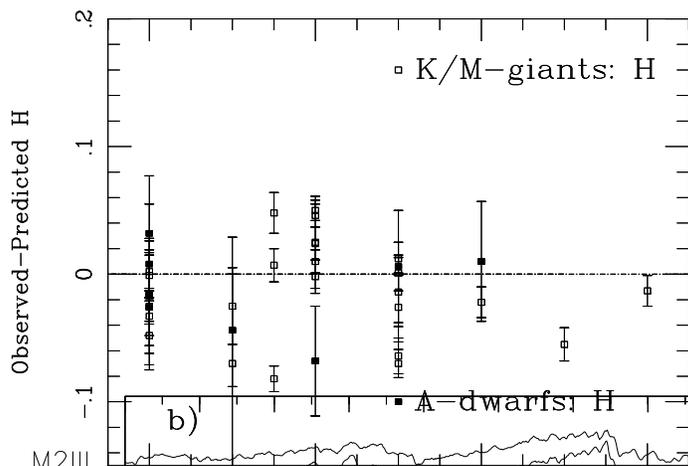
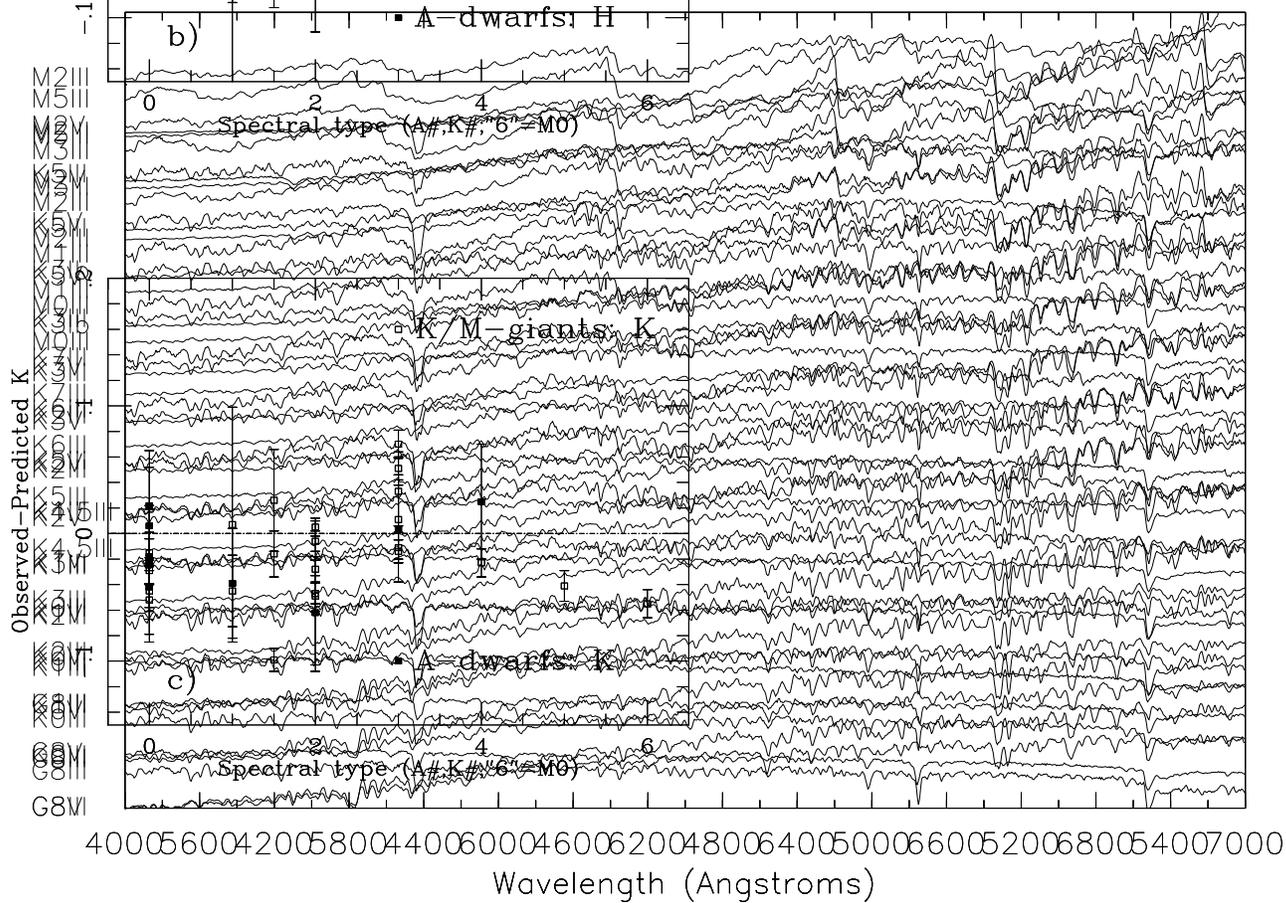
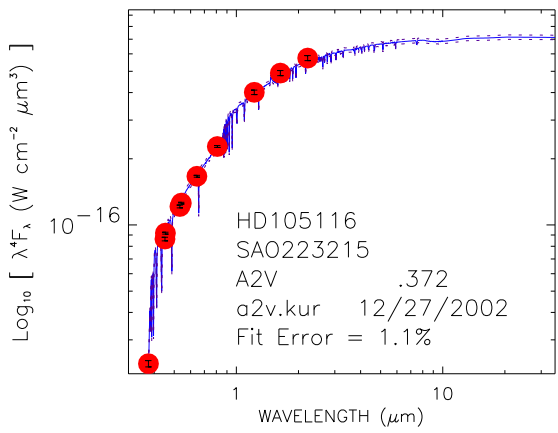
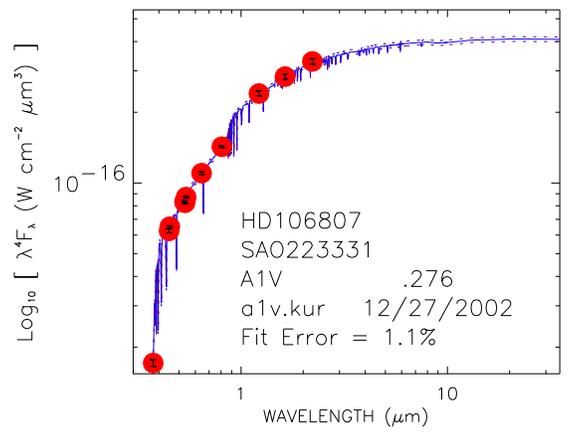
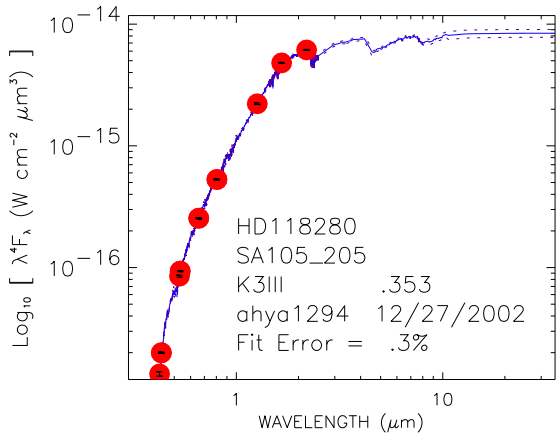
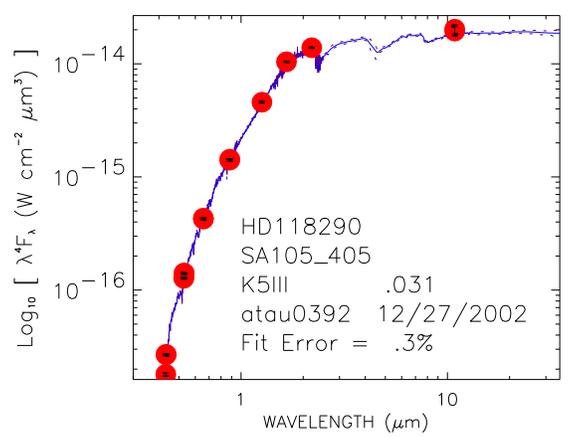
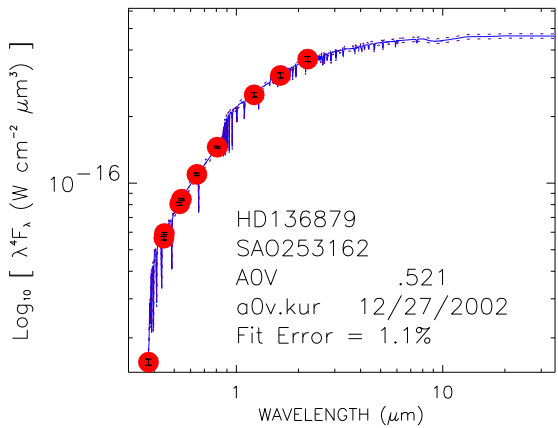
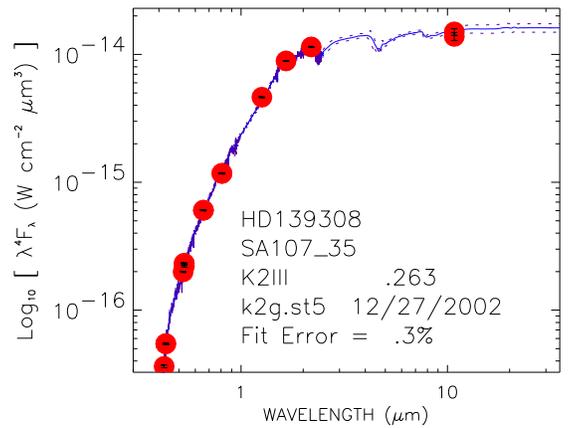
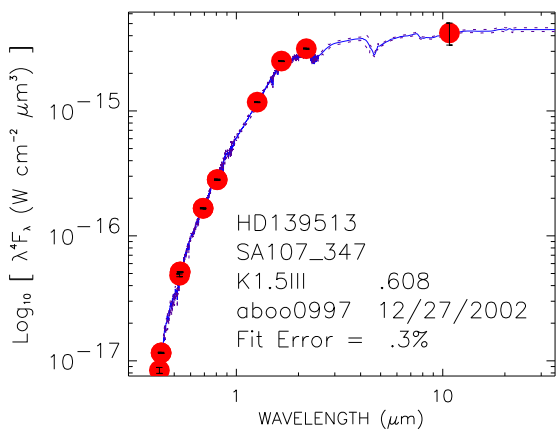
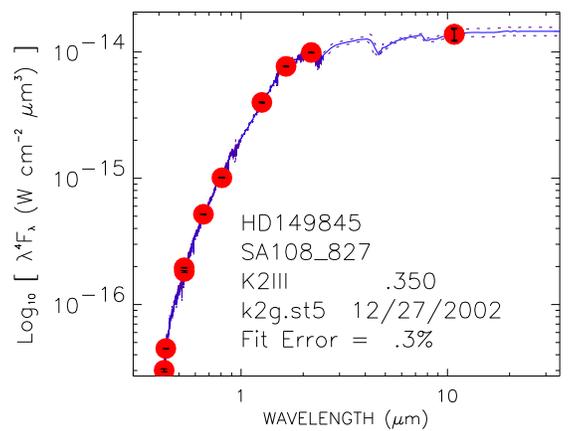
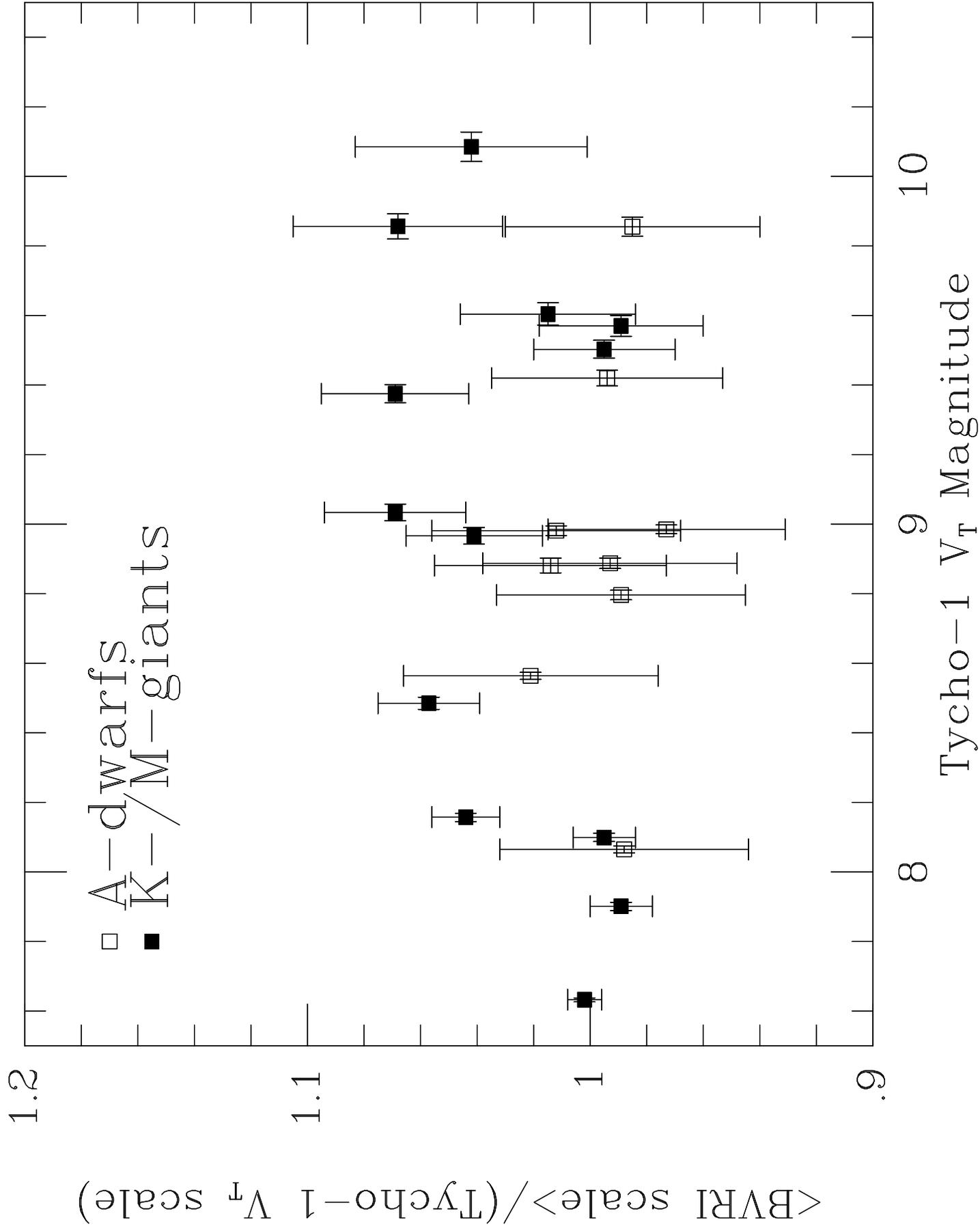
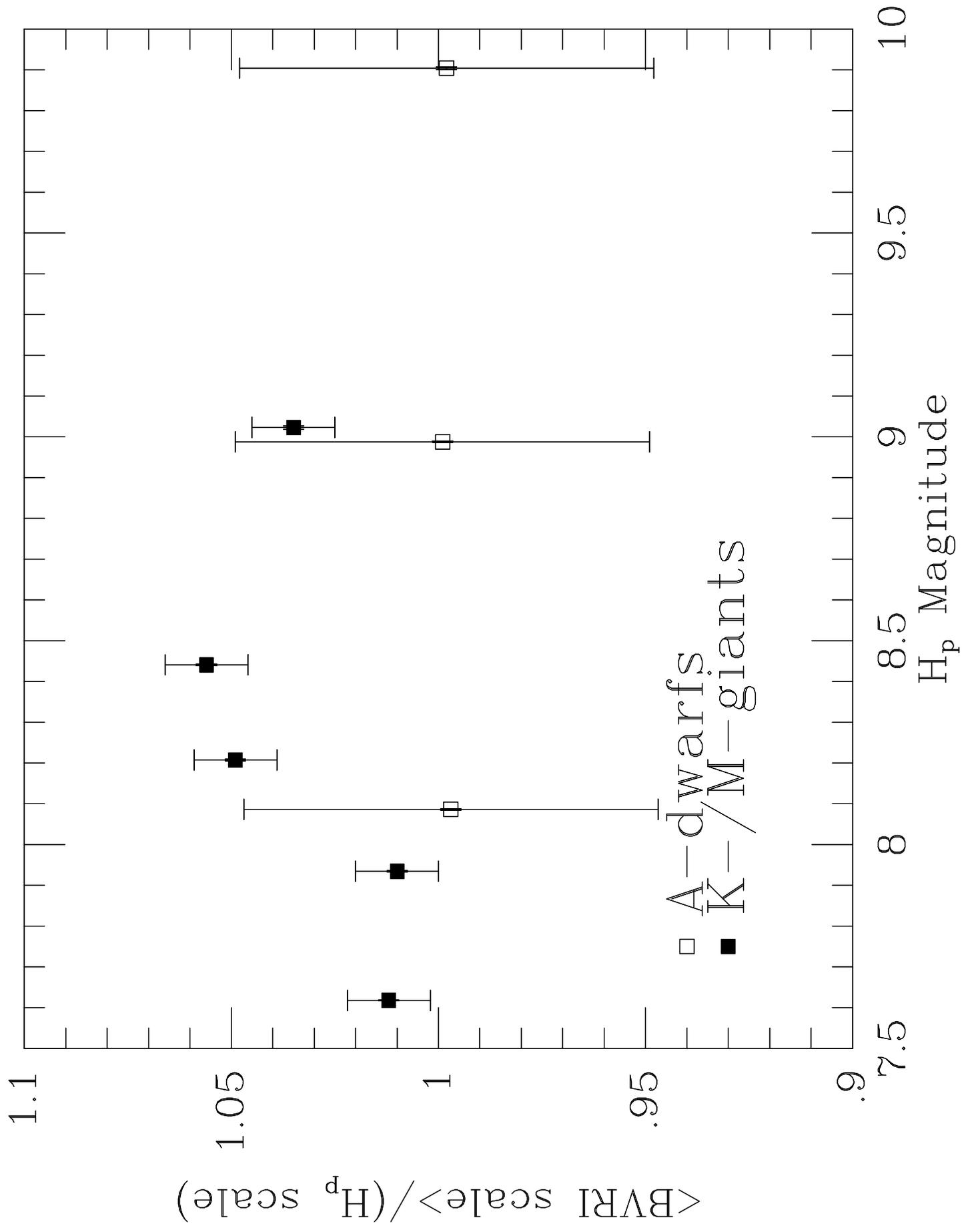
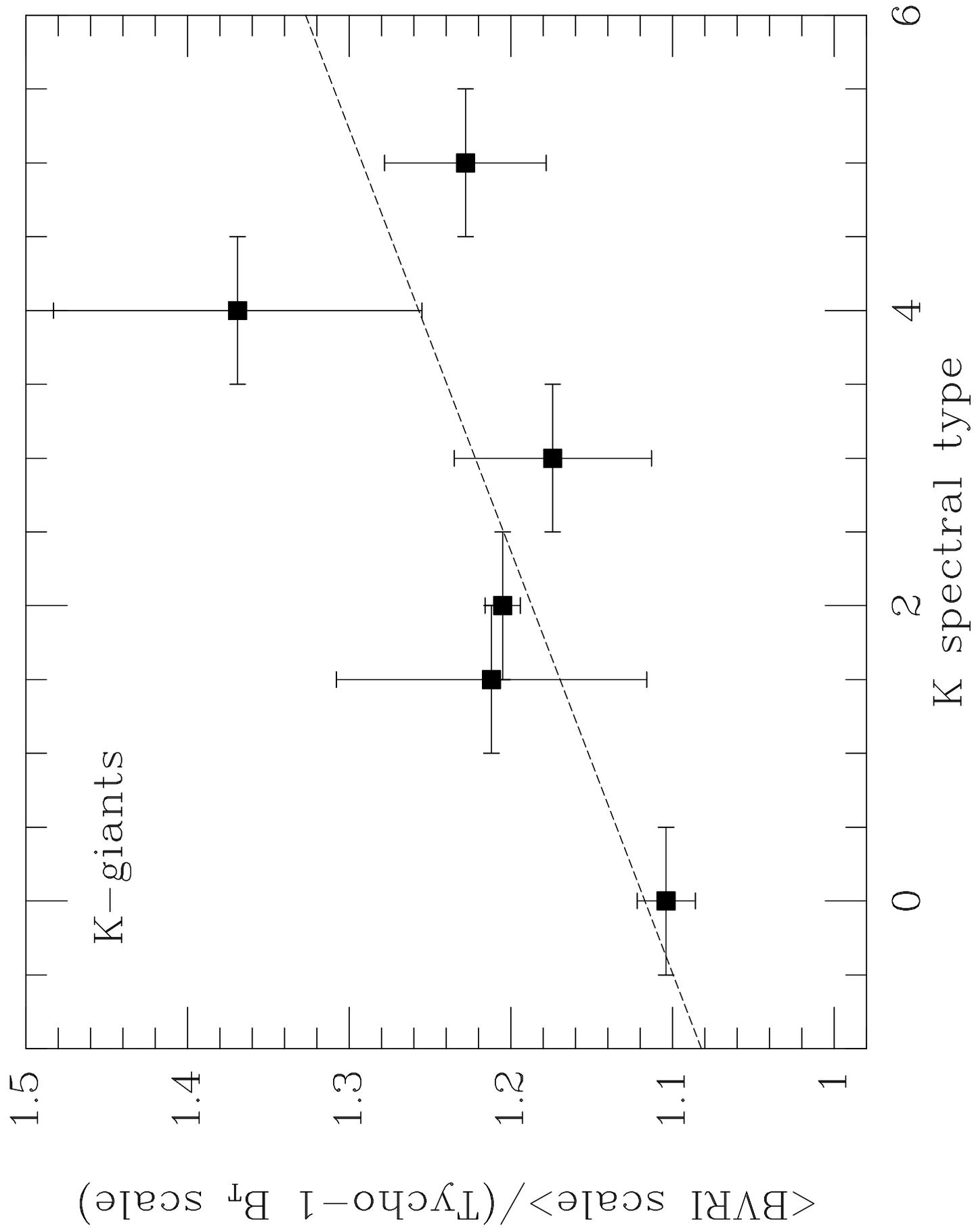
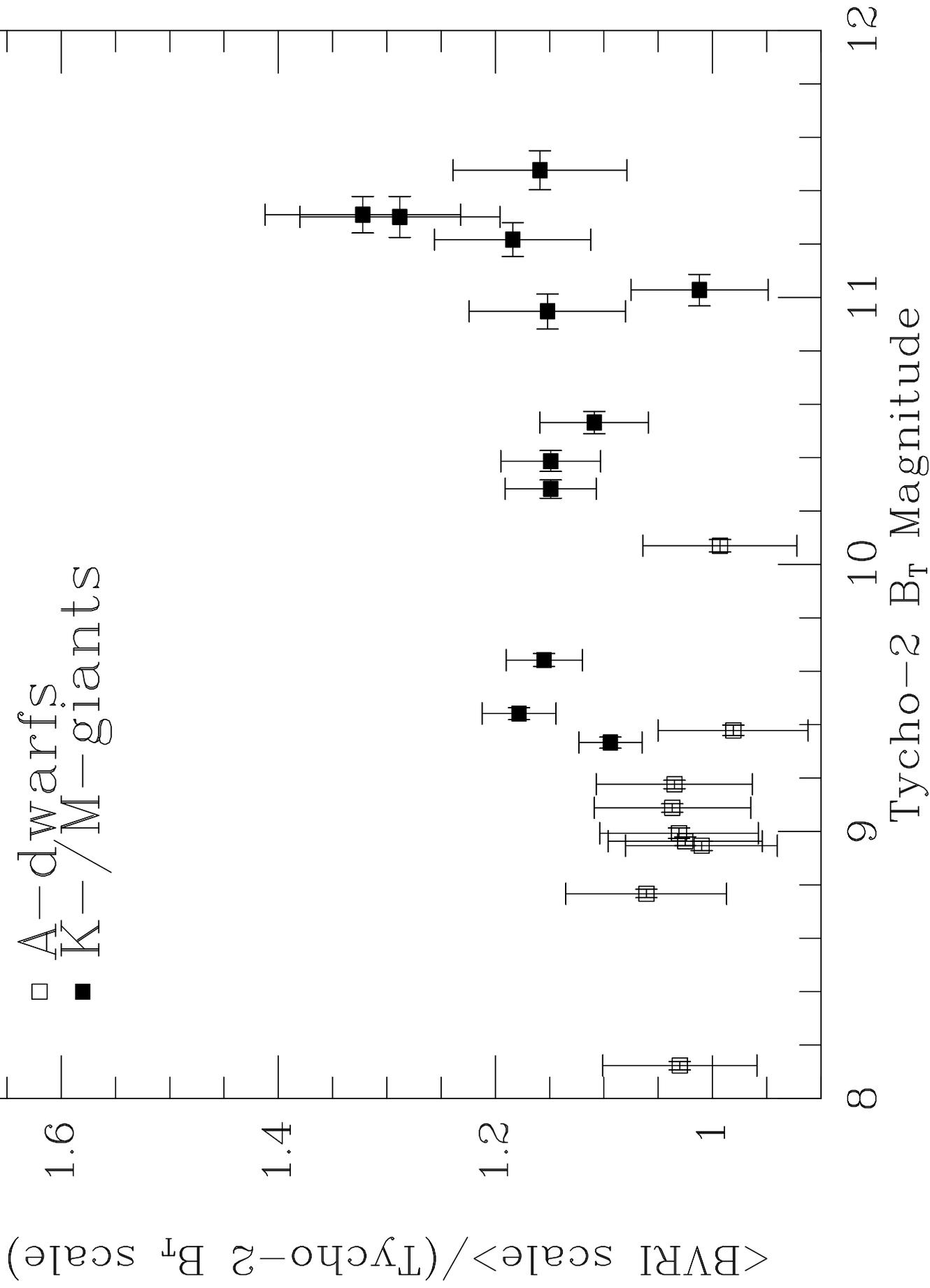
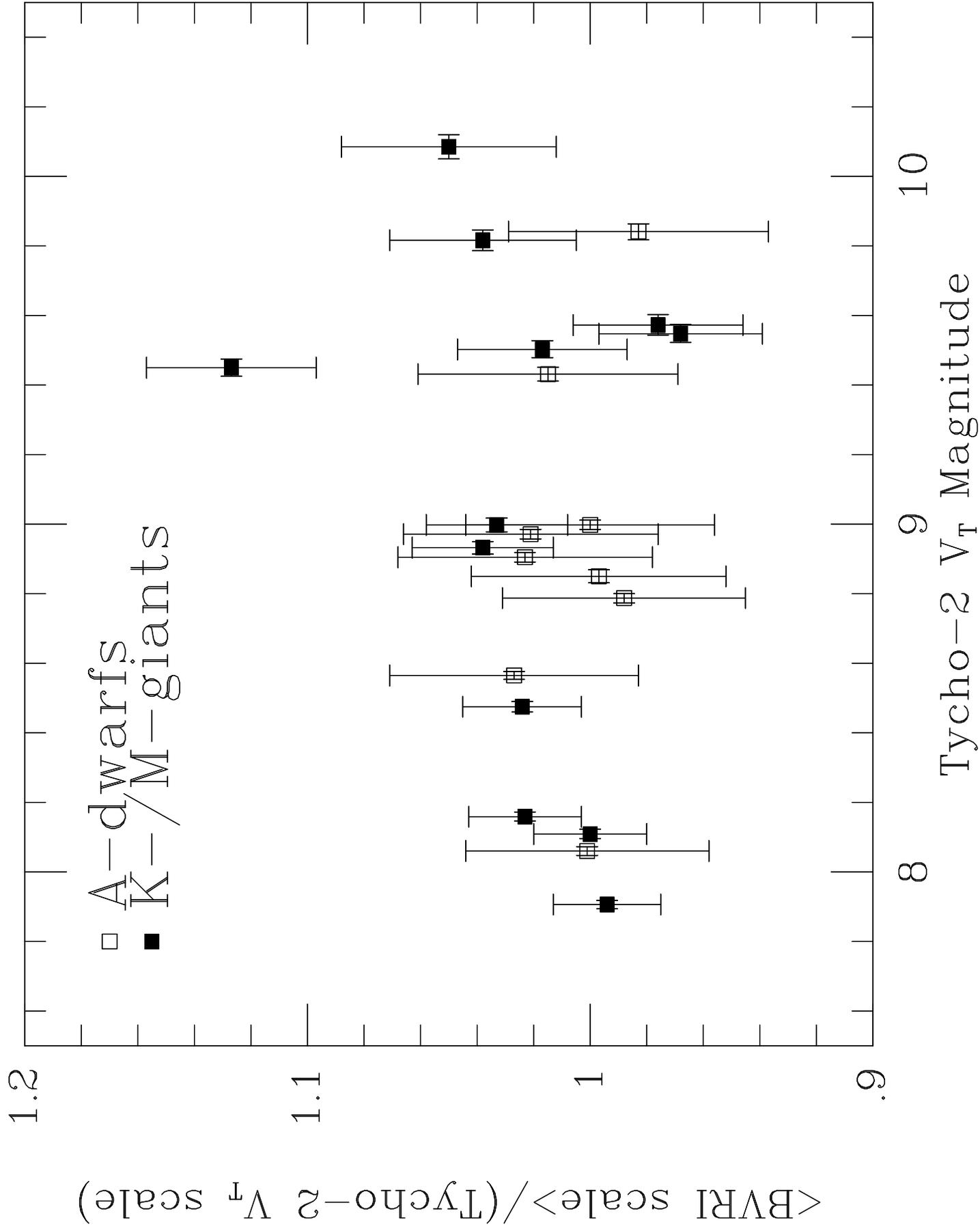
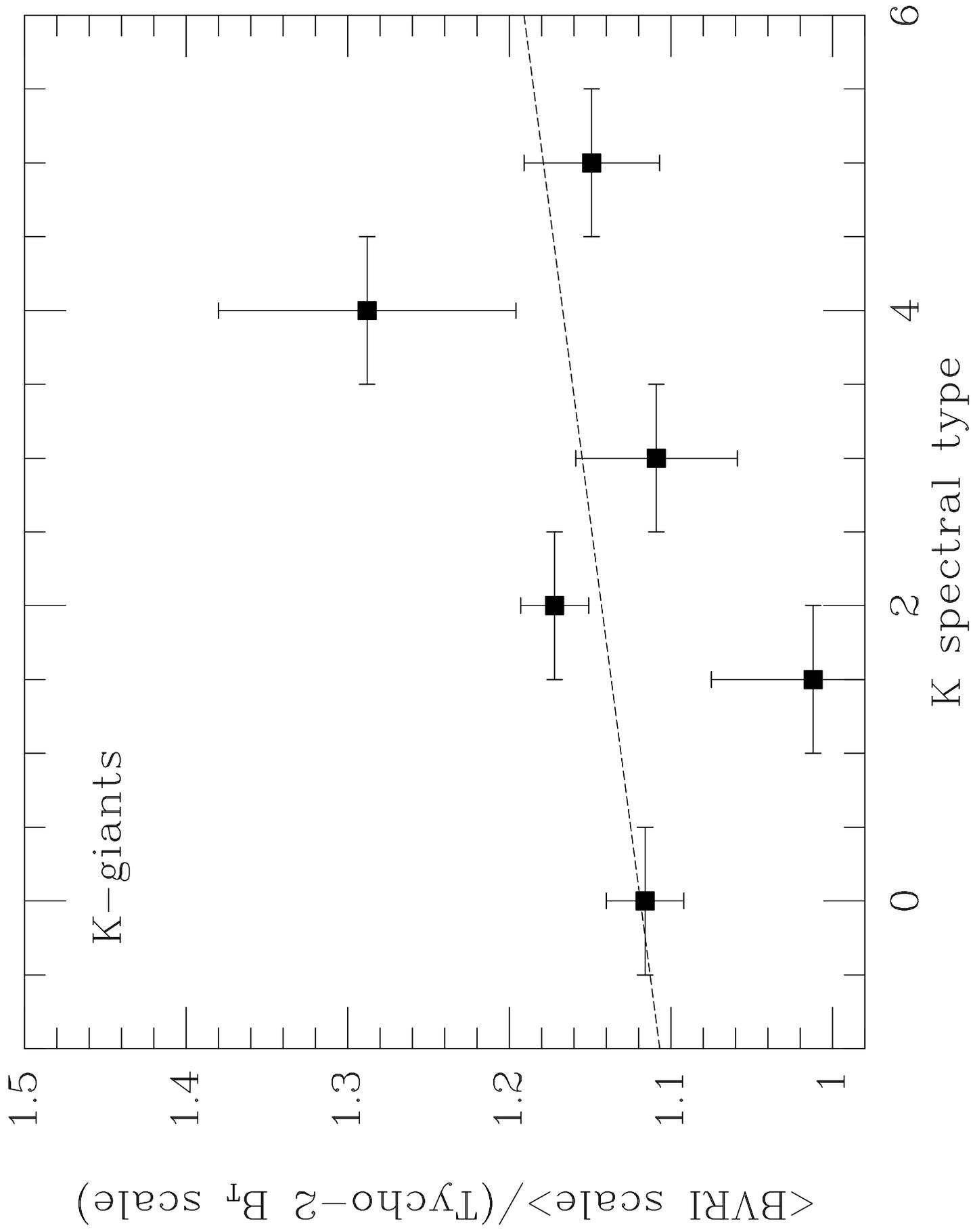
\end{document}